%% file: glambda_LMCS.tex
\documentclass{CSML}

\def\dOi{12(3:7)2016}
\lmcsheading%
{\dOi}
{1--39}
{}
{}
{Nov.~27, 2015}
{Sep.~\phantom06, 2016}
{}

\ACMCCS{[{\bf Theory of computation}]:  Semantics and
  reasoning---Program constructs\,/\,Program semantics\,/\,Program reasoning; [{\bf Software and its
    engineering}] Software notations and tools---General programming languages---Language features}

\usepackage{hyperref}

\usepackage{amsrefs}
\usepackage[all]{xy}
\SelectTips{cm}{} 
\usepackage{tikz-cd}
\usepackage{amssymb}
\usepackage{mathpartir}
\usepackage{stmaryrd}

\usepackage{todonotes}

\usepackage{listings}
\usepackage{lstautogobble}
\usepackage{comment}
\excludecomment{cameraversion}
\includecomment{arxivversion}

\theoremstyle{plain}

\begin{document}

\title[Guarded Lambda-Calculus]{The Guarded Lambda-Calculus \\
\footnotesize Programming and Reasoning with Guarded Recursion for Coinductive Types\rsuper*}

\author[R.~Clouston]{Ranald Clouston}	
\address{Department of Computer Science, Aarhus University, Denmark}	
\email{\{ranald.clouston, abizjak, bhugge, birkedal\}@cs.au.dk}  

\author[A.~Bizjak]{Ale\v{s} Bizjak}	
\address{\vspace{-18 pt}}	

\author[H.~Bugge Grathwohl]{Hans Bugge Grathwohl}	
\address{\vspace{-18 pt}}	

\author[L.~Birkedal]{Lars Birkedal}	
\address{\vspace{-18 pt}}	



\titlecomment{{\lsuper*}This is a revised and extended version of a FoSSaCS 2015 conference
paper~\cite{Clouston:Programming}.}



\begin{abstract}
  \noindent   We present the guarded lambda-calculus, an extension of the simply typed
  lambda-calculus with guarded recursive and coinductive types. The use of guarded
  recursive types ensures the productivity of well-typed programs. Guarded recursive
  types may be transformed into coinductive types by a type-former inspired by modal
  logic and Atkey-McBride clock quantification, allowing the typing of acausal functions.
  We give a call-by-name operational semantics for the calculus, and define adequate
  denotational semantics in the topos of trees. The adequacy proof entails that the
  evaluation of a program always terminates. We introduce a program logic with L{\"o}b
  induction for reasoning about the contextual equivalence of programs. We demonstrate
  the expressiveness of the calculus by showing the definability of solutions to Rutten's
  behavioural differential equations.
\end{abstract}

\maketitle


\input{macros}

\section*{Introduction}\enlargethispage{1 pt}

The problem of ensuring that functions on coinductive types are well-defined has
prompted a wide variety of work into productivity checking, and rule formats for
coalgebra.
\emph{Guarded recursion}~\cite{Coquand:Infinite} guarantees unique solutions for
definitions, as well as their \emph{productivity} -- any finite prefix of the solution can
be produced in finite time by unfolding --
by requiring that recursive calls on a coinductive data type be
nested under its constructor; for example, cons (written $\consin$) for
streams. This can sometimes be established by a simple syntactic check, as for
the stream $\mathsf{toggle}$ and binary stream function $\interleave$ below:
\begin{verbatim}
  toggle = 1 :: 0 :: toggle
  interleave (x :: xs) ys = x :: interleave ys xs
\end{verbatim}\nobreak
Such syntactic checks, however, exclude many valid definitions in the presence of higher
order functions. For example, consider the \emph{regular paperfolding sequence}
(also, more colourfully, known as the \emph{dragon curve
sequence}~\cite{Tabachnikov:Dragon}), which describes the sequence of left and right
folds induced by repeatedly folding a piece of paper in the same direction. This
sequence, with left and right folds encoded as $1$ and $0$, can be defined via the
function $\interleave$ as follows~\cite{Endrullis:Mix}:
\begin{verbatim}
  paperfolds = interleave toggle paperfolds
\end{verbatim}
This definition is productive, but the putative definition below, which also applies
$\interleave$ to two streams and so should apparently have the same type, is not:
\begin{verbatim}
  paperfolds' = interleave paperfolds' toggle
\end{verbatim}
This equation is satisfied by any stream whose \emph{tail} is the
regular paperfolding sequence, so lacks a unique solution. Unfortunately syntactic
productivity checking, such as that employed by the proof assistant
Coq~\cite{Coq:manual}, will fail to detect the difference between these programs, and
reject both.

A more flexible approach, first suggested by Nakano~\cite{Nakano:Modality}, is to
guarantee productivity via \emph{types}. A new modality, for which we follow Appel et
al.~\cite{Appel:Very} by writing $\LATER$ and using the name `later', allows us to
distinguish between data we have access to now, and data which we have only later.
This $\LATER$ must be used to guard self-reference in type definitions, so for example
\emph{guarded streams} over the natural numbers $\NAT$ are defined by the guarded
recursive equation
\[
  \gStream{\NAT} \defeq \NAT\times \LATER\gStream{\NAT}
\]
asserting that stream heads are available now, but tails only later.
The type of $\interleave$ will be $\gStream{\NAT}\to\LATER\gStream{\NAT}\to
\gStream{\NAT}$, capturing the fact the (head of the) first argument is needed
immediately, but the second argument is needed only later. In term definitions the
types of self-references will then be guarded by $\LATER$ also.
For example $\interleave\folds'\mathsf{toggle}$ becomes ill-formed, as the
$\folds'$ self-reference has type $\LATER\gStream{\NAT}$, rather than
$\gStream{\NAT}$ as required, but $\interleave\mathsf{toggle}\,\folds$ will be
well-formed. 

Adding $\LATER$ alone to the simply typed $\lambda$-calculus enforces a discipline
more rigid than productivity. For example the obviously productive stream function
\begin{verbatim}
  every2nd (x :: x' :: xs) = x :: every2nd xs
\end{verbatim}
cannot be typed because it violates \emph{causality}~\cite{Krishnaswami:Ultrametric}:
elements of the result stream depend on deeper elements of the argument stream. In
some settings, such as functional reactive programming, this is a desirable property, but
for productivity guarantees alone it is too restrictive -- we need the ability to remove
$\LATER$ in a controlled way. This is provided by the \emph{clock quantifiers}
of Atkey and McBride~\cite{Atkey:Productive}, which assert that all data is available
now. This does not trivialise the guardedness requirements because there
are side-conditions restricting how clock quantifiers may be introduced. Moreover
clock quantifiers allow us to recover first-class \emph{coinductive} types from guarded
recursive types, while retaining our productivity guarantees.

Note on this point that our presentation departs from Atkey and
McBride's~\cite{Atkey:Productive} by regarding the
`everything now' operator as a unary type-former, written $\blacksquare$ and called
`constant', rather than a quantifier. Observing that the types $\blacksquare A\to A$ and
$\blacksquare A\to\blacksquare\blacksquare A$ are always inhabited allows us to see
this type-former, via the Curry-Howard isomorphism, as an \emph{S4} modality, and
hence base this part of our calculus on the established typed calculi for intuitionistic
S4 (IS4) of Bierman and de Paiva~\cite{Bierman:Intuitionistic}. We will discuss the
trade-offs involved in this alternative presentation in our discussion of related work in
Section~\ref{sec:related}.

\paragraph{\emph{\textbf{Overview of our contributions.}}} In
Section~\ref{sec:calculus} we present the guarded $\lambda$-calculus, more briefly
referred to as the $\lambdanext$-calculus, extending the simply typed
$\lambda$-calculus
with guarded recursive and coinductive types. We define call-by-name operational
semantics, which will prevent the indefinite unfolding of recursive functions, an
obvious source of non-termination.
In Section~\ref{sec:denot} we define denotational semantics in the topos of
trees~\cite{Birkedal-et-al:topos-of-trees} which are \emph{adequate}, in the sense
that denotationally equal terms behave identically in any context, and as a corollary to
the logical relations argument used to establish adequacy, prove normalisation of the
calculus.

We are interested not only in \emph{programming} with guarded recursive and coinductive types, but also in \emph{proving} properties of these programs;
in Section~\ref{sec:logic} we show how the internal logic of the topos of trees induces
the program logic $\logiclambdanext$ for reasoning about the denotations of
$\lambdanext$-programs.
Given the adequacy of our semantics, this logic permits proofs about the operational
behaviour of terms. In Section~\ref{sec:definable-functions} we demonstrate the
expressiveness of the $\glambda$-calculus by showing the definability of
solutions to Rutten's behavioural differential
equations~\cite{Rutten:2003:bde}, and show that $\logiclambdanext$ can be used
to reason about them, as an alternative to standard bisimulation-based arguments.
In Section~\ref{sec:conc} we conclude with a discussion of related and further work.

\paragraph{This paper is based on a previously published conference
paper~\cite{Clouston:Programming}, but has been significantly revised and extended.}
We have improved the presentation of our results and examples
throughout the paper, but draw particular attention to the following changes:
\begin{itemize}
  \item
We present in the body of this paper many proof details that previously appeared only in
an appendix to the technical report version of the conference
paper~\cite{ARXIVVERSION}.
  \item
We discuss sums, and in particular the interaction between sums and the constant modality via the $\BOXSUM$ term-former, which previously appeared only in an
appendix to the technical report. We further improve on that discussion by presenting
conatural numbers as a motivating example; by giving new equational rules for
$\BOXSUM$ in Section~\ref{sec:logic_props}; and by proving a property of $\BOXSUM$
in Section~\ref{sec:logic-exs}.
  \item
We present new examples in Example~\ref{ex:awkward_programs} which show that
converting a program to type-check in the $\glambda$-calculus is not always
straightforward.
  \item
We give a more intuitive introduction to the logic $\logiclambdanext$ in
Section~\ref{sec:logic}, aimed at readers who are not experts in topos theory. In
particular we see how the guarded conatural numbers define the type of propositions.
  \item
We present new equational rules in Section~\ref{sec:logic_props} that reveal how
the explicit substitutions of the $\glambda$-calculus interact with real substitutions.
  \item
We present (slightly improved) results regarding total and inhabited types in the
$\glambda$-calculus in Section~\ref{sec:logic_props} which previously appeared only
in an appendix to the technical report. Relatedly, we have generalised the proof in
Example~\ref{ex:logic_exs}.\ref{ex:logic_exs_map} to remove its requirement that the
type in question is total and inhabited, by including a new equational rule regarding
composition for applicative functors.
  \item
We present formal results regarding behavioural differential equations in
Section~\ref{sec:definable-functions} which previously appeared only in an appendix to
the technical report.
  \item
We conduct a much expanded discussion of related and further work in
Section~\ref{sec:conc}.
\end{itemize}

\paragraph{We have implemented the $\lambdanext$-calculus in Agda, a process we found
  helpful when fine-tuning the design of our calculus.}  The implementation, with many
examples, is available
online.\footnote{\url{http://users-cs.au.dk/hbugge/bin/glambda.zip}}


\section{The Guarded Lambda-Calculus}\label{sec:calculus}

This section presents the guarded $\lambda$-calculus, more briefly referred to as the
$\glambda$-calculus, its call-by-name operational semantics, and its types, then gives
some examples.

\subsection{Untyped Terms and Operational Semantics}\label{sec:untyped}

In this subsection we will see the untyped $\glambda$-calculus and its call-by-name
operational semantics. This calculus takes the usual $\lambda$-calculus with natural
numbers, products, coproducts, and (iso-)recursion, and makes two extensions. First, the
characteristic operations of \emph{applicative functors}~\cite{McBride:Applicative}, here
called $\NEXT$ and $\APP$, are added, which will support the definition of causal
guarded recursive functions. Second, a $\PREV$ (previous) term-former is added,
inverse to $\NEXT$, that along with $\BOX$ and $\UNBOX$ term-formers
will support the definition of acausal functions without sacrificing guarantees of
productivity.

The novel term-formers of the $\glambda$-calculus are most naturally understood as
operations on its novel types. We will therefore postpone any examples of
$\glambda$-calculus terms until after we have seen its types.

Note that we will later add one more term-former, called $\BOXSUM$, to allow us to
write more programs involving the interaction of binary sums and the $\BOX$
term-former. We postpone discussion of this term-former until
Section~\ref{subsec:boxplus} to allow a cleaner presentation of the core system.

\begin{defi}\label{def:terms}
  \emph{Untyped $\glambda$-terms} are defined by the grammar
  \[
    \begin{array}{rcll}
      t & \bnfeq & x & \mbox{(variables)} \\
      &|&  \ZERO ~|~ \SUCC t & \mbox{(natural numbers)} \\
      &|& \UNIT ~|~ \langle t,t \rangle ~|~ \pi_1 t ~|~ \pi_2 t & \mbox{(products)} \\
      &|& \ABORT t ~|~ \IN_1 t ~|~ \IN_2 t ~|~ \CASE t \OF x_1. t; x_2. t &
        \mbox{(sums)} \\
      &|& \lambda x . t ~|~ tt & \mbox{(functions)} \\
      &|& \FOLD t ~|~ \UNFOLD t & \mbox{(recursion operations)} \\
      &|&   \NEXT t ~|~ \PREV \sigma.t ~|~ t \APP t & \mbox{(`later' operations)} \\
      &|& \BOX \sigma.t ~|~\UNBOX t & \mbox{(`constant' operations)}
    \end{array}
  \]
  where $\sigma$ is an \emph{explicit substitution}: a list of variables and terms
  $ [x_1 \explsubst t_1,\ldots,x_n\explsubst t_n]$, often abbreviated as $[\vec{x}
  \explsubst\vec{t}\,]$. We write $\PREV \iota. t$ for $\PREV[\vec{x}\explsubst
  \vec{x}].t$, where $\vec{x}$ is a list of all free variables of $t$, and write $\PREV t$
  where $\vec{x}$ is empty. We similarly write $\BOX\iota.t$ and $\BOX t$.

  The terms $\PREV[\vec{x}\explsubst\vec{t}\,].t$ and
  $\BOX [\vec{x}\explsubst\vec{t}\,].t$ bind all
  variables of $\vec{x}$ in $t$, but \emph{not} in $\vec{t}$. We adopt the convention that
  $\PREV$ and $\BOX$ have highest precedence.
\end{defi}

\begin{defi}\label{def:redrule}
The \emph{reduction rules} on closed $\glambda$-terms are
\[
  \begin{array}{rcll}
    \pi_d \langle t_1, t_2 \rangle & \red & t_d & \quad\mbox{\emph{($d\in\{1,2\}$)}} \\
    \CASE \IN_d t \OF x_1 . t_1; x_2 . t_2 & \red & t_d[t/x_d]
      & \quad\mbox{\emph{($d\in\{1,2\}$)}} \\
    (\lambda x . t_1) t_2 & \red & t_1 [t_2/x] \\
    \UNFOLD \FOLD t & \red & t \\
    \PREV [\vec{x} \explsubst \vec{t}\,].t & \red & \PREV (t[\vec{t}/\vec{x}])
      & \quad\mbox{\emph{($\vec{x}$ non-empty)}} \\
    \PREV\NEXT t & \red & t \\
    \NEXT t_1\APP\NEXT t_2 & \red & \NEXT(t_1 t_2) \\
    \UNBOX(\BOX [\vec{x} \explsubst \vec{t}\,].t) & \red & t[\vec{t}/\vec{x}] \\
  \end{array} 
\]
\end{defi}\medskip

\noindent All rules above except that concerning $\APP$ look like standard $\beta$-reduction,
removing `roundabouts' of introduction then elimination. A partial exception to this
observation are the $\PREV$ and $\NEXT$ rules; an apparently more conventional
$\beta$-rule for these term-formers would be
\begin{equation}\label{eq:general_prev_rule}
  \PREV [\vec{x} \explsubst \vec{t}\,].(\NEXT t) \; \red \; t[\vec{t}/\vec{x}]
\end{equation}
Where $\vec{x}$ is non-empty this rule might require us to reduce an \emph{open}
term to derive $\NEXT t$, for the computation to continue. But it is, as usual, easy to
construct examples of open terms that get stuck without reducing to a value, even where
they are well-typed (by the rules of the next subsection). Therefore a closed well-typed
term of form $\PREV [\vec{x} \explsubst \vec{t}\,].u$ may not see $u$ reduce to some
$\NEXT u'$, and so if equation \eqref{eq:general_prev_rule} were the only applicable
rule the term as a whole would also be stuck.

This is not necessarily a problem for us, because we are not interested in unrestricted
reduction. Such reduction is not compatible in a total calculus with the presence of
infinite structures such as streams, as we could choose
to unfold a stream indefinitely and hence normalisation would be lost.
In this paper we will instead adopt a strategy where we prohibit the reduction of open
terms; specifically we will use call-by-name evaluation.
In the case above we manage this by first applying the explicit substitution
without eliminating $\PREV$.

The rule involving $\APP$ is not a true $\beta$-rule, as $\APP$ is neither
introduction nor elimination, but is necessary to enable function application under a
$\NEXT$ and hence allow, for example, manipulation of the tail of a stream. It
corresponds to the `homomorphism' equality for applicative
functors~\cite{McBride:Applicative}.

We next impose our call-by-name strategy on these reductions.

\begin{defi}\label{def:value}
\emph{Values} are terms of the form
\[
  \SUCC^n\ZERO ~|~ \UNIT ~|~  \langle t,t\rangle ~|~ \IN_1 t ~|~ \IN_2 t ~|~
  \lambda x.t ~|~ \FOLD t ~|~ \NEXT t ~|~ \BOX \sigma.t
\]
where $\SUCC^n$ is a list of zero or more $\SUCC$ operators, and $t$ is any term.
\end{defi}

\begin{defi}\label{def:eval_ctx}
  \emph{Evaluation contexts} are defined by the grammar
  \[
    \begin{array}{rcl}
      E &\bnfeq& \cdot ~|~
      \SUCC E ~|~ \pi_1 E ~|~ \pi_2 E ~|~ \CASE E\OF x_1 . t_1; x_2 . t_2 ~|~ E t ~|~
      \UNFOLD E \\
      & | & \PREV E ~|~ E \APP t ~|~ v \APP E ~|~ \UNBOX E
    \end{array} 
  \]
\end{defi}\medskip

\noindent If we regard $\APP$ naively as function application, it is surprising in a call-by-name
setting that its right-hand side may be reduced. However both sides must be reduced
until they have main connective $\NEXT$, before the reduction rule for $\APP$ may be
applied. Thus the order of reductions of $\glambda$-terms cannot be identified with
the order of the call-by-name reductions of the corresponding $\lambda$-calculus term
with the novel connectives erased.

\begin{defi}
  \emph{Call-by-name reduction} has format $E[t]\red E[u]$, where $t\red u$ is a
  reduction rule. From now the symbol $\red$ will be reserved to refer to
  call-by-name reduction. We use $\redrt$ for the reflexive transitive closure of $\red$.
\end{defi}

Note that the call-by-name reduction relation $\red$ is deterministic.

\subsection{Types}

We now meet the typing rules of the $\glambda$-calculus, the most important feature of
which is the restriction of the fixed point constructor $\mu$ to \emph{guarded}
occurrences of recursion variables.

\begin{defi}\label{def:types}
  Open \emph{$\glambda$-types} are defined by the grammar
  \[
    \begin{array}{rcll}
      A & \bnfeq & \alpha & \mbox{(type variables)} \\
      &|&  \NAT & \mbox{(natural numbers)} \\
      &|& \ONE ~|~ A\times A & \mbox{(products)} \\
      &|& \EMPTY ~|~ A+A & \mbox{(sums)} \\
      &|& A\to A & \mbox{(functions)} \\
      &|& \mu\alpha.A & \mbox{(iso-recursive types)} \\
      &|&  \LATER A & \mbox{(later)} \\
      &|& \blacksquare A & \mbox{(constant)}
    \end{array}
  \]
  Type formation rules are defined inductively by the rules of Figure~\ref{fig:types}.
  In this figure $\nabla$ is a finite set of type variables, and
  a variable $\alpha$ is \emph{guarded in} a type $A$ if all
  occurrences of $\alpha$ are beneath an occurrence of $\LATER$ in the syntax tree.
 We adopt the convention that unary type-formers bind closer than binary type-formers.
  All types in this paper will be understood as closed unless explicitly stated otherwise.
\end{defi}

\begin{figure}
  \begin{mathpar}
    \inferrule*[right={$\alpha\in\nabla$}]{ }{\nabla \vdash \alpha}
    \and
    \inferrule*{ }{\nabla \vdash \NAT}
    \and
    \inferrule*{ }{\nabla \vdash \ONE}
    \and
    \inferrule*{%
      \nabla \vdash A_1 \\
      \nabla \vdash A_2}{%
      \nabla\vdash A_1\times A_2}
    \and
    \inferrule*{ }{\nabla \vdash \EMPTY}
    \and
    \inferrule*{%
      \nabla \vdash A_1 \\
      \nabla \vdash A_2}{%
      \nabla\vdash A_1+A_2}
    \and
    \inferrule*{%
      \nabla \vdash A_1 \\
      \nabla \vdash A_2}{%
      \nabla\vdash A_1\to A_2}
    \and
    \inferrule*[right={$\alpha\,\mathsf{guarded\,in}\,A$}]{%
      \nabla,\alpha \vdash A}{%
      \nabla \vdash \mu\alpha.A}
    \and
    \inferrule*{%
      \nabla \vdash A}{%
      \nabla\vdash \LATER A}
    \and
    \inferrule*{%
      \cdot \vdash A}{%
      \nabla\vdash \blacksquare A}
  \end{mathpar}
  \caption{Type formation for the $\glambda$-calculus}
  \label{fig:types}
\end{figure}

Note that the guardedness side-condition on the $\mu$ type-former and the prohibition
on the formation of $\blacksquare A$ for open $A$ together create a prohibition on
applying $\mu\alpha$ to any $\alpha$ with $\blacksquare$ above it, for example
$\mu\alpha.\blacksquare\LATER\alpha$ or $\mu\alpha.\LATER\blacksquare\alpha$. This
accords with our intuition that fixed points will exist only where a recursion variable is
`displaced in time' by a $\LATER$. The constant type-former $\blacksquare$ destroys
any such displacement by giving `everything now'.

\begin{defi}\label{def:typing}
The \emph{typing judgments} are given in Figure~\ref{fig:typing}. There 
$\Gamma$ is a \emph{typing context}, i.e. a finite set of variables $x$, each associated
with a type $A$, written $x:A$.
In the side-conditions to the $\PREV$ and $\BOX$
rules, types are \emph{constant} if all occurrences of $\LATER$ are beneath an
occurrence of $\blacksquare$ in their syntax tree.
\end{defi}

\begin{figure}
  \begin{mathpar}
    \inferrule*{ }{\Gamma, x : A \vdash x : A}
    \and
    \inferrule*{ }{\Gamma \vdash \ZERO:\NAT}
    \and
    \inferrule*{\Gamma \vdash t:\NAT}{%
      \Gamma \vdash \SUCC t:\NAT}
    \and
    \inferrule*{ }{\Gamma \vdash \UNIT : \ONE}
    \and
    \inferrule*{\Gamma \vdash t_1 : A \\%
      \Gamma \vdash t_2 : B}{%
      \Gamma \vdash \langle t_1,t_2 \rangle : A \times B}
    \and
    \inferrule*{\Gamma \vdash t: A \times B}{%
      \Gamma \vdash \pi_1 t : A}
    \and
    \inferrule*{\Gamma \vdash t: A \times B}{%
      \Gamma \vdash \pi_2 t : B}
    \and
    \inferrule*{\Gamma \vdash t:\EMPTY}{%
      \Gamma \vdash \ABORT t:A}
    \and
    \inferrule*{\Gamma \vdash t: A}{%
      \Gamma \vdash \IN_1 t : A + B}
    \and
    \inferrule*{\Gamma \vdash t: B}{%
      \Gamma \vdash \IN_2 t : A + B}
    \and
    \inferrule*{\Gamma \vdash t: A+B \\
      \Gamma,x_1:A\vdash t_1:C \\
      \Gamma,x_2:B\vdash t_2:C}{%
      \Gamma \vdash \CASE t \OF x_1 . t_1; x_2 . t_2:C}
    \and
    \inferrule*{\Gamma, x : A \vdash t : B}{%
      \Gamma \vdash \lambda x . t : A \to B}
    \and
    \inferrule*{\Gamma \vdash t_1 : A \to B \\%
      \Gamma \vdash t_2 : A}{%
      \Gamma \vdash t_1 t_2 : B}
    \and
    \inferrule*{\Gamma \vdash t:A[\mu\alpha.A/\alpha]}{%
      \Gamma \vdash \FOLD t : \mu\alpha.A}
    \and
    \inferrule*{\Gamma \vdash t : \mu\alpha.A}{%
      \Gamma \vdash \UNFOLD t : A[\mu\alpha.A/\alpha]}
   \and
   \inferrule*{\Gamma \vdash t : A}{%
     \Gamma \vdash \NEXT t : \LATER A}
    \and
     \inferrule*[right={$A_1,\ldots,A_n\,\mathsf{constant}$}]{%
      x_1:A_1,\ldots,x_n:A_n \vdash t:\LATER A \\
      \Gamma\vdash t_1:A_1 \\
      \cdots \\
      \Gamma\vdash t_n:A_n }{%
      \Gamma \vdash \PREV [x_1\explsubst t_1,\ldots,x_n\explsubst t_n].t : A}
    \and
    \inferrule*{\Gamma \vdash t_1 : \LATER (A \to B) \\%
      \Gamma \vdash t_2 : \LATER A}{%
      \Gamma \vdash t_1 \APP t_2 : \LATER B}
    \and
    \inferrule*[right={$A_1,\ldots,A_n\,\mathsf{constant}$}]{%
      x_1:A_1,\ldots,x_n:A_n \vdash t:A \\
      \Gamma\vdash t_1:A_1 \\
      \cdots \\
      \Gamma\vdash t_n:A_n }{%
      \Gamma \vdash \BOX [x_1\explsubst t_1,\ldots,x_n\explsubst t_n].t :\blacksquare A}
    \and
    \inferrule*{\Gamma \vdash t:\blacksquare A}{%
      \Gamma \vdash \UNBOX t: A}
  \end{mathpar}
  \caption{Typing rules for the $\glambda$-calculus}
  \label{fig:typing}
\end{figure}

The \emph{constant} types exist `all at once', due to the absence of $\LATER$
or presence of $\blacksquare$; this condition corresponds to the freeness of the
clock variable in Atkey and McBride~\cite{Atkey:Productive} (recalling that this paper's
work corresponds to the use of only one clock). Its use as a side-condition to
$\blacksquare$-introduction in Figure~\ref{fig:typing} recalls (but is more
general than) the `essentially modal' condition in the natural deduction calculus of
Prawitz~\cite{Prawitz:Natural} for the modal logic Intuitionistic S4 (IS4). The term
calculus for IS4 of Bierman and de Paiva~\cite{Bierman:Intuitionistic},
on which this calculus is most closely based, uses the still more restrictive requirement
that $\blacksquare$ be the main connective. 
This would preclude some functions that
seem desirable, such as the isomorphism $\lambda n.\BOX \iota.n:\NAT\to\blacksquare
\NAT$.

The presence of explicit substitutions attached to the $\PREV$ and $\BOX$ can seem
heavy notationally, but in practice the burden on the programmer seems quite
small, as in all examples we will see, $\PREV$ appears only in its syntactic sugar forms
\[
  \inferrule*[right={$A_1,\ldots,A_n\,\mathsf{constant}$}]{%
    x_1:A_1,\ldots,x_n:A_n \vdash t:\LATER A }{%
    \Gamma,x_1:A_1,\ldots,x_n:A_n \vdash \PREV \iota. t: A}
  \qquad
  \inferrule*{%
    \cdot\vdash t : \LATER A}{%
    \Gamma \vdash \PREV t : A}
\]
and similarly for $\BOX$. One might therefore ask why the more general form involving
explicit substitutions is necessary. The answer is that the `sugared' definitions above are
not closed under substitution: we need $(\PREV \iota. t)[\vec{u}/\vec{x}] = \PREV
[\vec{x}\explsubst\vec{u}].t$. In general getting substitution right in the presence of
side-conditions can be rather delicate. The solution we use, namely \emph{closing} the
term $t$ to which $\PREV$ (or $\BOX$) is applied to protect its variables, comes
directly from Bierman and de Paiva's calculus for IS4~\cite{Bierman:Intuitionistic}; see
this reference for more in-depth discussion of the issue, and in particular how a failure to
account for this issue causes problems for the calculus of Prawitz~\cite{Prawitz:Natural}.
Similar side-conditions have also caused problems in the closely related area of calculi
with clocks -- see the identification by Bizjak and M{\o}gelberg~\cite{Bizjak:Model} of a
problem with the type theory presented in earlier work by
M{\o}gelberg~\cite{Mogelberg:tt-productive-coprogramming}.

\begin{lem}[Subject Reduction for Closed Terms]
$\vdash t:A$ and $t\redrt u$ implies $\vdash u:A$.
\qed
\end{lem}

Note that the reduction rule
\[
\PREV [\vec{x} \explsubst \vec{t}\,].t \,\red\, \PREV (t[\vec{t}/\vec{x}])
\]
plainly violates subject reduction for open terms: the right hand side is only
well-defined if $t[\vec{t}/\vec{x}]$ has no free variables, because the explicit
substitution attached to $\PREV$ must close all open variables.

\subsection{Examples}

We may now present example $\glambda$-programs and their typings. We will first give
causal programs without use of the constant modality $\blacksquare$, then show
how this modality expands the expressivity of the language, and finally show two
examples of productive functions which are a bit trickier to fit within our language.

\begin{exa}\label{ex:programs}\hfill
\begin{enumerate}
\item
  The type of guarded recursive streams over some type $A$, written $\gStream{A}$, is,
  as noted in the introduction, defined as $\mu\alpha.A\times\LATER\alpha$. Other
  guarded recursive types can be defined, such as infinite binary trees as
  $\mu\alpha.A\times\LATER(\alpha\times\alpha)$, conatural numbers $\gCoNat$ as
  $\mu\alpha.1+\LATER\alpha$, and colists as $\mu\alpha.1+(A\times\LATER\alpha)$.
  We will focus on streams in this section, and look more at $\gCoNat$ in
  Section~\ref{subsec:boxplus}.
\item
  We define guarded versions of the standard stream functions cons (written infix as
  $\consin$), head, and tail as obvious:
  \[
    \begin{array}{rclcl}
    \consin &\defeq& \lambda x.\lambda s.\FOLD\langle x,s\rangle &:&
      A\to\LATER\gStream{A}\to\gStream{A} \\
    \head &\defeq& \lambda s.\pi_1\UNFOLD s &:& \gStream{A}\to A \\
    \tail &\defeq& \lambda s.\pi_2\UNFOLD s &:& \gStream{A}\to\LATER\gStream{A}
    \end{array}
  \]
  We can then use the $\APP$
  term-former to make observations deeper into the stream:
  \[
    \begin{array}{rclcl}
    \gsecond &\defeq&
      \lambda s.(\NEXT\head)\APP(\tail s) &:& \gStream{A}\to\LATER A \\
    \gthird &\defeq&
      \lambda s.(\NEXT\gsecond)\APP(\tail s) &:& \gStream{A}\to\LATER\LATER A
      \;\cdots
    \end{array}
  \]
\item
  To define guarded recursive functions we need a fixed point combinator.
  Abel and Vezzosi~\cite{Abel:Formalized} gave a guarded version of Curry's $Y$
  combinator in a similar calculus; for variety we present a version of Turing's fixed
  point combinator.

  Recall from the standard construction that if we had a $\mu$ type-former with no
  guardedness requirements, then a
  combinator $\Theta$ with type $(A\to A)\to A$ could be defined, for any type $A$, by
  the following:
  \[
  \begin{array}{rclcl}
    \Rec_A &\defeq& \mu\alpha.(\alpha\to(A\to A)\to A) \\
    \theta &\defeq& \lambda y.\lambda f.f((\UNFOLD y)yf) &:& \Rec_A \to (A \to A) \to A \\
    \Theta &\defeq& \theta  (\FOLD \theta) &:& (A\to A)\to A
  \end{array}
  \]  
  To see that $\Theta$ does indeed behave as a fixpoint, note
  that $\Theta f$ unfolds in one step to $f((\UNFOLD\FOLD\theta)(\FOLD\theta)f)$. But
  $\UNFOLD\FOLD$ eliminates%
  \footnote{With respect to call-by-name evaluation this program's next reduction will
  depend on the shape of $f$, but it is enough for this discussion to see that
  $\UNFOLD\FOLD\theta$ is equal to $\theta$ in the underlying equational theory.}%
  , so we have $f(\Theta f)$.

  What then is the guarded version of this combinator? Following the need for the
  recursion variable to be guarded, and the original observation of
  Nakano~\cite{Nakano:Modality}
  that guarded fixed point combinators should have type $(\LATER A\to A)\to A$, we
  reconstruct the type $\Rec_A$ by the addition of later modalities in the appropriate
  places. The terms $\theta$
  and $\Theta$ can then be constructed by adding $\NEXT$ term-formers, and replacing
  function application with $\APP$, to the original terms so that they type-check:
  \[
  \begin{array}{rcl}
    \Rec_A &\defeq& \mu\alpha.(\LATER\alpha\to(\LATER A\to A)\to A) \\
    \theta &\defeq& \lambda y.\lambda f.f((\NEXT \lambda z.\UNFOLD z)\APP y \APP
       \NEXT y \APP \NEXT f) \,:\\ && \LATER \Rec_A \to (\LATER A \to A) \to A \\
    \Theta &\defeq& \theta  (\NEXT \FOLD \theta) \,:\, (\LATER A\to A)\to A
  \end{array}
  \]
  The addition of these novel term-formers is fairly mechanical; the only awkward point
  comes when we cannot unfold $y$ directly because it has type $\LATER\Rec_A$
  rather than $\Rec_A$, so we must introduce the expression $\lambda z.\UNFOLD z$.

  Now $\Theta f$ reduces to
  \[
    f((\NEXT \lambda z.\UNFOLD z)\APP (\NEXT\FOLD\theta) \APP (\NEXT\NEXT\FOLD\theta) \APP \NEXT f)
  \] 
  But the reduction rule for $\APP$ allows us to take $\NEXT$ out the front and replace
  $\APP$ by normal application:
  \[
    f(\NEXT((\lambda z.\UNFOLD z)(\FOLD\theta)(\NEXT\FOLD\theta)f))
  \] 
  Applying the $\lambda$-expression and eliminating $\UNFOLD\FOLD$ yields
  $f(\NEXT\Theta f)$. In other words, we have defined a standard fixed point except that
  a $\NEXT$ is added to the term to record that the next application of the fixed point
  combinator must take place one step in the future. We will be able to be more formal
  about this property of $\Theta$ in Lemma~\ref{prop:theta-is-a-fp}, once we have
  introduced the program logic $\logiclambdanext$ for reasoning about
  $\glambda$-programs.

  Note that the inhabited type $(\LATER A\to A)\to A$ does not imply that all types are
  inhabited, as there is not in general a function $\LATER A \to A$. This differs from the
  standard presentation of fixed point combinators that leads to inconsistency.
\item\label{ex:zeros}
  Given our fixed point combinator we may now build some guarded streams; for
  example, the simple program (in pseudocode)
  \begin{lstlisting}
    zeros = 0 :: zeros
  \end{lstlisting}
  is captured by the term
  \[
    \mathsf{zeros} \,\defeq\, \Theta \lambda s.(\ZERO \consin s)
  \]
  of type $\gStream{\NAT}$. Here $s$ has type $\LATER\gStream{\NAT}$, and so the
  function that the fixed point is applied to has type $\LATER\gStream{\NAT}\to
  \gStream{\NAT}$; exactly the type expected by $\Theta$.

  Note however that the plainly unproductive stream definition
  \begin{lstlisting}
  circular = circular
  \end{lstlisting}
  cannot be defined within this calculus, although it is it apparently definable via a
  standard fixed point combinator as $\Theta \lambda s.s$; in our calculus the type of
  the recursion variable $s$ must be preceded by a $\LATER$ modality.
\item
  For a slightly more sophisticated example, consider the standard map function on
  streams:
  \[
    \map \,\defeq\, \lambda f.\Theta\lambda m.\lambda s.(f\head s)\consin(m\APP\tail s) \,:\, (A\to B)\to \gStream{A}\to \gStream{B}
  \]
   Here the recursion variable $m$ has type $\LATER(\gStream{A}\to \gStream{B})$.
  \label{ex:map}
\item
  We can define two more standard stream functions -- $\iterate$, which takes a
  function $A\to A$ and a head $A$, and produces a stream by applying the
  function repeatedly, and $\interleave$, which interleaves two streams -- in the obvious
  ways:
  \[
    \begin{array}{rclcl}
      \iterate' &\defeq& \lambda f.\Theta \lambda g.\lambda x.x\consin(g\APP\NEXT(fx))
        &:& (A\to A) \to A \to \gStream{A} \\
      \interleave' &\defeq& \Theta\lambda g.\lambda s.\lambda t.(\head s)\consin
        (g\APP (\NEXT t)\APP\tail s) &:& \gStream{A} \to \gStream{A} \to \gStream{A}
    \end{array}
  \]
  These definitions are correct but are less informative than they could be, as they do
  not record the temporal aspects of these functions, namely that
  (in the case of $\iterate$) the function, and (in the case of $\interleave$) the second
  stream, are not used until the next time step. We could alternatively use the definitions
  \[
    \begin{array}{rclcl}
      \iterate &\defeq& \lambda f.\Theta \lambda g.\lambda x.x\consin
        (g\APP(f\APP\NEXT x)) &:& \LATER(A\to A) \to A \to \gStream{A} \\
      \interleave &\defeq& \Theta\lambda g.\lambda s.\lambda t.(\head s)\consin
        (g\APP t\APP\NEXT\tail s) &:& \gStream{A} \to \LATER\gStream{A} \to \gStream{A}
    \end{array}
  \]
  These definitions are in fact more general:
  \[
    \begin{array}{rcl}
      \iterate' f\, x &=& \iterate (\NEXT f)\, x \\
      \interleave' s\, t &=& \interleave s\, (\NEXT t)
    \end{array}
  \]
  Indeed the example of the regular paperfolding sequence from the introduction
  shows that the more general and informative version can also be more useful:
  \[
    \begin{array}{rclcl}
      \tyrol &\defeq& \Theta\lambda s.(\SUCC\ZERO)\consin
        (\NEXT(\ZERO\consin s)) &:& \gStream{\NAT} \\
      \folds &\defeq& \Theta\lambda s.\interleave\tyrol\,s &:& \gStream{\NAT}
     \end{array}
  \]
  The recursion variable $s$ in $\folds$ has type $\LATER\gStream{\NAT}$, which
  means it cannot be given as the second argument to $\interleave'$ -- only the more
  general $\interleave$ will do.
  However the erroneous definition of the regular paperfolding sequence that replaced
  $\interleave\tyrol\,s$ with $\interleave' s\,\tyrol$ cannot be typed.

  Another example of a function that (rightly) cannot be typed in $g\lambda$ is
  a $\mathsf{filter}$ function on streams which eliminates elements that fail some
  boolean test; as all elements may fail the test, the function is not productive.
  \label{exa:iterate_and_interleave}
\item
  $\mu$-types define \emph{unique} fixed points, carrying both initial algebra and final
  coalgebra structure. For example, the type $\gStream{A}$ is both the initial algebra and
  the final coalgebra for the functor $A\times\LATER\mbox{-}$.
  This contrasts with the usual case of streams, which are merely the final coalgebra for the
  functor $A\times\mbox{-}$; the initial algebra for this functor is trivial.
  To see the dual structure of guarded recursive types, consider the functions%
  \footnote{These are usually called $\mathsf{fold}$ and $\mathsf{unfold}$; we avoid
  this because of the name clash with our term-formers.}%
  \[
    \begin{array}{rclcl}
    \mathsf{initial} &\hspace{-0.4em}\defeq&\hspace{-0.4em} \Theta\lambda g.\lambda f.\lambda s.f
      \langle\head s,g\APP\NEXT f\APP \tail s\rangle
      &\hspace{-0.4em}:&\hspace{-0.4em}
      ((A\times\LATER B)\to B)\to \gStream{A}\to B \\
    \mathsf{final} &\hspace{-0.4em}\defeq&\hspace{-0.4em} \Theta\lambda g.\lambda f.\lambda x.
      (\pi_1 (f x))\consin(g\APP\NEXT f\APP \pi_2 (f x))
      &\hspace{-0.4em}:&\hspace{-0.4em}
      (B\to A\times\LATER B)\to B\to \gStream{A}
    \end{array}
    \]
  For example, $\mathsf{\map}\,h:\gStream{A}\to\gStream{A}$ can be written as
  $\mathsf{initial}\,\lambda x.(h (\pi_1 x))\consin(\pi_2 x)$, or as
  $\mathsf{final}\,\lambda s.\langle h(\head s),\tail s\rangle$.
\end{enumerate}
\end{exa}\medskip

\noindent The next examples involve the $\PREV$ (previous) term-former
and the constant modality $\blacksquare$.\newpage

\begin{exa}\label{exa:constant_progs}\hfill
\begin{enumerate}
\item
  The $\blacksquare$ type-former lifts guarded recursive streams to coinductive
  streams, as we will make precise in Example~\ref{ex:denote_streams}. We
  define $\Stream{A} \defeq \blacksquare\gStream{A}$. We can then define
  versions of cons, head, and tail operators for coinductive streams:
  \[
    \begin{array}{rclcl}
    \cons &\defeq& \lambda x.\lambda s.\BOX\iota.x\consin(\UNBOX s)
      &:& A\to\Stream{A}\to\Stream{A} \\
    \limhead &\defeq& \lambda s . \head (\UNBOX s) &:& \Stream{A}\to A \\
    \limtail &\defeq& \lambda s . \BOX \iota . \PREV \iota . \tail (\UNBOX s)
      &:& \Stream{A}\to\Stream{A}
    \end{array}
  \]
  Note that $\cons$ is well-defined only if $A$ is a constant type.
  Note also that we must `unbox' our coinductive stream
  to turn it into a guarded stream before we operate on it. This explains why we retain
  our productivity guarantees. Finally, note
  the absence of $\LATER$ in the types. Indeed we can define observations
  deeper into the stream with no hint of later, for example
  \[
    \mathsf{2nd} \,\defeq\, \lambda s.\limhead (\limtail s) \,:\, \Stream{A} \to A
  \]
\item
  We have a general way to lift boxed functions to functions on boxed types, via
  the `limit' function
  \[
    \limit \,\defeq\, \lambda f.\lambda x.\BOX \iota.(\UNBOX f)(\UNBOX x) \,:\,
      \blacksquare(A\to B)\to\blacksquare A\to\blacksquare B
  \]
  This allows us to lift our guarded stream functions from
  Example~\ref{ex:programs} to coinductive stream functions, provided that the function
  in question is defined in a constant environment. For example
  \[
    \constmap \defeq \lambda f.\limit\BOX\iota.(\map f):
      (A\to B)\to\Stream{A}\to\Stream{B}
  \]
  is definable if $A\to B$ is a constant type (which is to say, $A$ and $B$ are constant
  types).
  \label{exa:lim}
\item
  The more sophisticated acausal function $\everysecond:\Stream{A}\to\gStream{A}$ is
  \[
    \Theta\lambda g.\lambda s.(\limhead s)\consin (g\APP\NEXT(\limtail(\limtail s)))
  \]
  Note that it takes a \emph{coinductive} stream $\Stream{A}$ as argument.
  The function with coinductive result type is then
  $\lambda s.\BOX \iota.\everysecond s:\Stream{A}\to\Stream{A}$.
  \label{exa:every2nd}
\item
  Guarded streams do not define a monad, as the standard `diagonal' join function
  $\gStream{(\gStream{A})}\to\gStream{A}$ cannot be defined, as for example the
  second element of the second stream in $\gStream{(\gStream{A})}$ has type
  $\LATER\LATER A$, while the second element of the result stream should have type
  $\LATER A$ -- the same problem as for $\everysecond$ above. However we can define
  \[
    \mathsf{diag} \,\defeq\, \Theta\lambda f.(\limhead(\limhead s))\consin
      (f\APP\NEXT(\limtail(\limtail s))) \,:\,
      \Stream{(\Stream{A})}\to\gStream{A}
  \]
  The standard join function is then $\lambda s.\BOX\iota.\mathsf{diag}\,s:
  \Stream{(\Stream{A})}\to\Stream{A}$.
\end{enumerate}
\end{exa}\medskip

\noindent In the examples above the construction of typed $\glambda$-terms from the standard
definitions of productive functions required little ingenuity; one merely applies the new
type- and term-formers in the `necessary places' until everything type-checks. This
appears to be the case with the vast majority of such functions. However, below are
two counter-examples, both from Endullis et al.~\cite{Endrullis:Circular}, where a bit
more thought is required:\newpage

\begin{exa}\label{ex:awkward_programs}\hfill
\begin{enumerate}
\item
  The \emph{Thue-Morse sequence} is a stream of booleans which can be defined (in
  pseudo-code) as
  \begin{lstlisting}
    thuemorse  = 0 :: tl (h thuemorse)
    h (0 :: s) = 0 :: 1 :: (h s)
    h (1 :: s) = 1 :: 0 :: (h s)
  \end{lstlisting}
  The definition of $\mathsf{thuemorse}$ is productive only because the helper stream
  function $\mathsf{h}$ produces two elements of its result stream after reading one
  element of its input stream. To see that this is crucial, observe that if we replace
  $\mathsf{h}$ by the identity stream function, $\mathsf{thuemorse}$ is no longer
  productive. The type of $\mathsf{h}$ therefore needs to be something
  other than $\gStream{(\ONE+\ONE)}\to\gStream{(\ONE+\ONE)}$. But it does not have
  type $\LATER\gStream{(\ONE+\ONE)}\to\gStream{(\ONE+\ONE)}$
  because it needs to read the head of its input stream before it produces the first
  element of its output stream. Capturing this situation -- a stream function that produces
  nothing at step zero, but two elements at step one -- seems too fine-grained to fit
  well with our calculus with $\LATER$.

  The simplest solution is to modify the definition above by unfolding the definition
  of $\mathsf{thuemorse}$ once:
   \begin{lstlisting}
    thuemorse = 0 :: 1 :: h (tl (h thuemorse))
  \end{lstlisting} 
  This equivalent definition \emph{would} remain productive if we replaced $\mathsf{h}$
  with the identity, and so $\mathsf{h}$ can be typed
  $\gStream{(\ONE+\ONE)}\to\gStream{(\ONE+\ONE)}$ without problem.
\item
  The definition below of the \emph{Fibonacci word} is similar to the example above,
  but shows that the situation can be even more intricate:
  \begin{lstlisting}
    fibonacci  = 0 :: tl (f fibonacci)
    f (0 :: s) = 0 :: 1 :: (f s)
    f (1 :: s) = 0 :: (f s)
  \end{lstlisting}
  Here the helper function $\mathsf{f}$, if given a stream with head $0$, produces
  nothing at step zero, but two elements at step one, as for $\mathsf{h}$ above. But
  given a stream with head $1$, it produces only one element at step one. Therefore
  the erroneous definition
  \begin{lstlisting}
    fibonacci' = 1 :: tl (f fibonacci')
  \end{lstlisting}
  whose head is $1$ rather than $0$, is not productive. Productivity hence depends on an
  inspection of \emph{terms}, rather than merely types, in a manner clearly beyond the
  scope of our current work.
  
  Again, this can be fixed by unfolding the definition once:
  \begin{lstlisting}
    fibonacci = 0 :: 1 :: f (tl (f fibonacci))
  \end{lstlisting}
\end{enumerate}
\end{exa}

\subsection{Sums and the Constant Modality}\label{subsec:boxplus}

Atkey and McBride's calculus with clocks~\cite{Atkey:Productive} includes as a primitive
notion \emph{type equalities} regarding the interaction of clock quantification with other
type-formers. They note that most of these equalities are not essential, as in many cases
mutually inverse terms between the sides of the equalities are definable. However
this is not so with, among other cases, binary sums. Binary sums present a similar
problem for our calculus. We can define a term
\[
  \lambda x.\BOX \iota.\CASE x\OF x_1.\IN_1\UNBOX x_1;x_2.\IN_2\UNBOX x_2:
  (\blacksquare A+\blacksquare B)\to \blacksquare(A+B)
\]
in our calculus but no term in general in the other direction. Unfortunately such a
term is essential to defining some basic operations involving coinductive
types involving sums. For example we define the (guarded and coinductive) conatural
numbers as
\[
  \begin{array}{rcl}
    \gCoNat &\defeq& \mu\alpha.(1+\LATER\alpha) \\
    \CoNat &\defeq& \blacksquare\gCoNat
  \end{array}
\]
These correspond to natural numbers with infinity, with such programs definable upon
them as
\[
  \begin{array}{rclcl}
    \mathsf{cozero} &\defeq& \FOLD(\IN_1\UNIT) &:& \gCoNat \\
    \mathsf{cosucc} &\defeq& \lambda n.\FOLD(\IN_2(\NEXT n)) &:& \gCoNat\to\gCoNat \\
    \mathsf{infinity} &\defeq& \Theta\lambda n.\FOLD(\IN_2 n) &:& \gCoNat
  \end{array}
\]
As a guarded recursive construction, $\gCoNat$ defines a unique fixed point.
In particular its coalgebra map $\gpred$ (for `predecessor') is simply
\[
  \gpred \,\defeq\, \lambda n.\UNFOLD n \,:\, \gCoNat \to 1+\LATER\gCoNat
\]
Now the coinductive type $\CoNat$ should be a coalgebra also, so we should be able to
define a function $\pred:\CoNat\to 1+\CoNat$ similarly. However a term of type $\CoNat$
must be unboxed before it is unfolded, and the type $1+\LATER\gCoNat$ that results is
not constant, and so we cannot apply $\PREV$ and $\BOX$ to map from
$\LATER\gCoNat$ to $\CoNat$.

Our solution is to introduce a new term-former $\BOXSUM$ which will allow us to define
a term
\[
  \lambda x.\BOXSUM \iota.\UNBOX x:\blacksquare(A+B)\to\blacksquare A+\blacksquare B
\]

\begin{defi}[ref. Definitions~\ref{def:terms}, \ref{def:redrule}, \ref{def:eval_ctx},
\ref{def:typing}]\label{defi:boxsum}
We extend the grammar of $\glambda$-\emph{terms} by
\[
  \begin{array}{rcl}
    t & \bnfeq & \cdots ~|~ \BOXSUM \sigma.t
  \end{array}
\]
where $\sigma$ is an explicit substitution. We abbreviate terms with
$\BOXSUM$ as for $\PREV$ and $\BOX$.

We extend the reduction rules with
\[
  \begin{array}{rcll}
    \BOXSUM[\vec{x} \explsubst \vec{t}\,].t & \red & \BOXSUM t[\vec{t}/\vec{x}] &
      \quad\mbox{\emph{($\vec{x}$ non-empty)}} \\
    \BOXSUM\IN_d t & \red & \IN_d\BOX t & \quad\mbox{\emph{($d\in\{1,2\}$)}}
  \end{array} 
\]
  We do not change the definition of values of Definition~\ref{def:value}. We extend
  the definition of evaluation contexts with
  \[
    \begin{array}{rcl}
      E &\bnfeq& \cdots ~|~ \BOXSUM E
    \end{array}
  \]
  Finally, we add the new typing judgment
  \begin{mathpar}
    \inferrule*[right={$A_1,\ldots,A_n\,\mathsf{constant}$}]{%
      x_1:A_1,\ldots,x_n:A_n \vdash t:B_1+B_2 \\
      \Gamma\vdash t_1:A_1 \\
      \cdots \\
      \Gamma\vdash t_n:A_n }{%
      \Gamma \vdash \BOXSUM[x_1\explsubst t_1,\ldots,x_n\explsubst t_n].t:
        \blacksquare B_1+\blacksquare B_2}
  \end{mathpar}
\end{defi}\medskip

\noindent Returning to our example, we can define the term $\pred:\CoNat\to 1+\CoNat$ as
\[
  \lambda n.\CASE(\BOXSUM\iota.\UNFOLD\UNBOX n)\OF x_1.\IN_1\UNIT;
    x_2.\IN_2\BOX\iota.\PREV\iota.\UNBOX x_2
\]

\section{Denotational Semantics and Normalisation}\label{sec:denot}

This section gives denotational semantics for $\glambda$-types and terms, as
objects and arrows in the topos of trees~\cite{Birkedal-et-al:topos-of-trees}, the
presheaf category over the first infinite ordinal $\omega\defeq1\leq2\leq\cdots$%
\footnote{It would be more standard to start this pre-order at $0$, but we start at $1$ to
maintain harmony with some equivalent presentations of the topos of trees and related
categories which have a vacuous stage $0$; we shall see such a presentation in
Section~\ref{sec:sheaf}}
(we give a concrete definition below). The denotational semantics are shown to be sound
and, by a logical relations argument, adequate with respect to the operational semantics.
Normalisation follows as a corollary of this argument.

\subsection{The topos of trees}\label{sec:tot}

This section introduces the mathematical model in which our denotational semantics
will be defined.

\begin{defi}
The \emph{topos of trees} $\trees$ has, as objects $X$, families of sets $X_1,X_2,$
$\ldots$ indexed by the positive integers, equipped with families of \emph{restriction
functions} $\res{X}{i}:X_{i+1}\to X_i$ indexed similarly. Arrows $f:X\to Y$ are families
of functions $f_i:X_i\to Y_i$ indexed similarly obeying the \emph{naturality} condition
$f_i\circ\res{X}{i}=\res{Y}{i}\circ f_{i+1}$:
\[
  \xymatrix{
    X_1 \ar[d]_{f_1} & X_2 \ar[l]_-{\res{X}{1}} \ar[d]_{f_2} & X_3 \ar[l]_-{\res{X}{2}}
      \ar[d]_{f_3} & \cdots \ar[l]_-{\res{X}{3}} \\
    Y_1 & Y_2 \ar[l]^-{\res{Y}{1}} & Y_3  \ar[l]^-{\res{Y}{2}}  & \cdots
      \ar[l]^-{\res{Y}{3}}
  }
\]
\end{defi}\medskip

\noindent Given an object $X$ and positive integers $i \leq j$ we write $\restriction_i$ for the
function $X_j\to X_i$ defined by composing the restriction functions $\res{X}{k}$ for
$k\in \{i,i+1,\ldots,j-1\}$, or as the identity where $i=j$.

$\trees$ is a cartesian closed category with products and coproducts defined pointwise.
Note that by naturality it holds that for any arrow $f:X\to Y+Z$, positive integer $n$,
and element $x\in X_n$, $f_i\circ\restriction_i(x)$ must be an element of the same side of the
sum for all $i \leq n$. The exponential $A^B$ has, as its component sets $(A^B)_i$, the
set of $i$-tuples $(f_1:A_1\to B_1,\ldots,f_i:A_i\to B_i)$ obeying the naturality condition,
and projections as restriction functions.

\begin{defi}\label{def:functors}\hfill
\begin{enumerate}
\item
  The category of sets $\sets$ is a full subcategory of $\trees$ via the functor $\Delta:
  \sets\to\trees$ that maps sets $Z$ to the $\trees$-object
  \[
  \xymatrix{
    Z & Z \ar[l]_-{id_Z} & Z \ar[l]_-{id_Z} & \cdots \ar[l]_-{id_Z}
  }
  \]
  and maps functions $f$ by $(\Delta f)_i=f$ similarly.

  The full subcategory of \emph{constant objects} consists of $\trees$-objects which are \emph{isomorphic} to objects of the form $\Delta Z$.
  These are precisely the objects whose restriction functions are bijections.
  In particular the terminal object $1$ of $\trees$ is $\Delta \{\ast\}$, the initial object is $\Delta\emptyset$, and the \emph{natural numbers object} is $\Delta\mathbb{N}$;
  
  We will abuse notation slightly and treat constant objects as if they were actually of the form $\Delta Z$, i.e., if $X$ is constant and $x \in X_i$ we will write $x$ also, for example, for the element $\left(r_{i}^X\right)^{-1}(x) \in X_{i+1}$.
\item
  $\Delta$ is left adjoint to the `global elements' functor $hom_{\trees}(1,\mbox{--})$.
  We write $\blacksquare$ for the endofunctor $\Delta\circ hom_{\trees}(1,\mbox{-}):
  \trees\to\trees$. Then $\UNBOX:\blacksquare\natto id_{\trees}$ is the counit of the
  comonad associated with this adjunction. Concretely, for any $\trees$-object $X$ and
  $x\in hom_{\trees}(1,X)$ we have
  $\UNBOX_i(x)=x_i$, i.e. the $i$'th component of $x:1\to X$ applied to the unique
  element $\ast$:
  \[
  \xymatrix{
    hom_{\trees}(1,X) \ar[d]_-{x\,\mapsto\, x_1} & hom_{\trees}(1,X) \ar[l]_-{id}
      \ar[d]_-{x\,\mapsto\, x_2} & hom_{\trees}(1,X) \ar[l]_-{id}
      \ar[d]_-{x\,\mapsto\, x_3} & \cdots \ar[l]_-{id} \\
    X_1  & X_2 \ar[l]^-{\res{X}{1}} & X_3 \ar[l]^-{\res{X}{2}} & \cdots
      \ar[l]^-{\res{X}{3}}
  }
  \]
  The global elements functor can also be understood by considering an $\trees$-object
  $X$ as a diagram in $\sets$; then $hom_{\trees}(1,X)$ is its \emph{limit}, and so
  $\blacksquare X$ is this limit considered as a $\trees$-object.
  \label{def:functorsconstant}
\item
  $\LATER:\trees\to\trees$ is defined by mapping $\trees$-objects $X$ to
  \[
  \xymatrix{
    \{\ast\} & X_1 \ar[l]_-{!} & X_2 \ar[l]_-{\res{X}{1}} & \cdots \ar[l]_-{\res{X}{2}}
  }
  \]
  That is, $(\LATER X)_1=\{\ast\}$ and $(\LATER X)_{i+1}=X_i$, with
  $\res{\LATER X}{1}$ defined uniquely and $\res{\LATER X}{i+1}
  =\res{X}{i}$. The $\LATER$ functor acts on arrows $f:X\to Y$ by $(\LATER f)_1=
  id_{\{\ast\}}$ and $(\LATER f)_{i+1}=f_i$. The natural transformation
  $\NEXT:id_{\trees}\natto\LATER$ has, for each component $X$, $\NEXT_1$ uniquely
  defined and $\NEXT_{i+1}=\res{X}{i}$:
  \[
  \xymatrix{
    X_1 \ar[d]_{!} & X_2 \ar[l]_-{\res{X}{1}} \ar[d]_{\res{X}{1}} & X_3
      \ar[l]_-{\res{X}{2}} \ar[d]_{\res{X}{2}} & \cdots \ar[l]_-{\res{X}{3}} \\
    \{\ast\} & X_1 \ar[l]^-{!} & X_2  \ar[l]^-{\res{X}{1}}  & \cdots
      \ar[l]^-{\res{X}{2}}
  }
  \]
  \label{def:functorslater}
\end{enumerate}
\end{defi}

\subsection{Denotational Semantics}\label{sec:sound_denot}

We may now see how the $\glambda$-calculus can be interpreted soundly in the topos
of trees.

\begin{defi}\label{def:types_denote}
  We interpret types in context $\nabla\vdash A$, where $\nabla$ contains $n$ free
  variables, as functors $\den{\nabla\vdash A}:(\trees^{op}\times\trees)^n\to\trees$,
  usually written $\den{A}$. This mixed variance definition is necessary as variables may
  appear negatively or positively.
\begin{itemize}
\item
  $\den{\nabla,\alpha\vdash\alpha}$ is the projection of the objects or arrows
  corresponding to \emph{positive} occurrences of $\alpha$, e.g. $\den{\alpha}
  (\vec{W},X,Y)=Y$;
\item
  $\den{\NAT}$, $\den{\ONE}$, and $\den{\EMPTY}$ are the constant functors
  $\Delta\mathbb{N}$, $\Delta\{\ast\}$, and $\Delta\emptyset$ respectively;
\item
  $\den{A_1\times A_2}(\vec{W})=\den{A_1}(\vec{W})
  \times\den{A_2}(\vec{W})$. The definition of the functor on $\trees$-arrows is
  likewise pointwise;
\item
  $\den{A_1+A_2}(\vec{W})=\den{A_1}(\vec{W})+\den{A_2}(\vec{W})$ similarly;
\item
  $\den{\mu\alpha.A}(\vec{W}) = \mathsf{Fix}(F)$, where
  $F:(\trees^{op}\times\trees)\to\trees$
  is the functor given by $F(X,Y) = \den{A}(\vec{W},X,Y)$ and
  $\mathsf{Fix}(F)$ is the unique (up to isomorphism) $X$ such that
  $F(X,X)\cong X$. The existence of such $X$
  relies on $F$ being a suitably locally contractive functor,
  which follows by Birkedal et al.~\cite[Section~4.5]{Birkedal-et-al:topos-of-trees}
  and the fact that $\blacksquare$ is only ever applied to closed types.
  This restriction on $\blacksquare$ is necessary because the functor
  $\blacksquare$ is not \emph{strong}.
\item
  $\den{A_1\to A_2}(\vec{W})=\den{A_2}
  (\vec{W})^{\den{A_2}(\vec{W}')}$ where $\vec{W}'$ is $\vec{W}$
  with odd and even elements switched to reflect change in polarity, i.e. $(X_1,Y_1,
  \ldots)'=(Y_1,X_1,\ldots)$;
\item
  $\den{\LATER A},\den{\blacksquare A}$ are defined
  by composition with the functors $\LATER,\blacksquare$ (Def.~\ref{def:functors}).
\end{itemize}
\end{defi}

\begin{exa}\label{ex:denote_streams}\hfill
\begin{enumerate}
  \item\label{ex:gStream}
  $\den{\gStream{\NAT}}$ is the $\trees$-object
  \[
  \xymatrix{
    \mathbb{N} & \mathbb{N}\times\mathbb{N} \ar[l]_-{pr_1} &
      (\mathbb{N}\times\mathbb{N})\times\mathbb{N} \ar[l]_-{pr_1} & \cdots
      \ar[l]_-{pr_1}
  }
  \]
  where the $pr_1$ are first projection functions. This is intuitively the object of
  \emph{approximations} of streams -- first the head, then the first two elements, and so
  forth. Conversely,  
  $\den{\Stream{\NAT}}=\Delta(\mathbb{N}^{\omega})$, so it is
  the constant object of streams, as usually defined in $\sets$. This can also be
  understood as the limit of the approximations given by $\den{\gStream{\NAT}}$.

  More generally, any polynomial functor $F$ on $\sets$ can be assigned a
  $\glambda$-type $A_F$ with a free type variable $\alpha$ that occurs guarded. The
  denotation of $\blacksquare\mu\alpha. A_F$ will then be the constant object of the
  carrier of the final coalgebra for
  $F$~\cite[Theorem 2]{Mogelberg:tt-productive-coprogramming}. Therefore
  $\blacksquare$ is the modality that takes us from guarded recursive constructions to
  coinductive constructions.
  \item\label{ex:gCoNat}
  $\den{\gCoNat}$ is the $\trees$-object
  \[
  \xymatrix{
    2 & 3 \ar[l]_-{\res{\Omega}{1}} & 4 \ar[l]_-{\res{\Omega}{2}} & \cdots
    \ar[l]_-{\res{\Omega}{3}}
  }
  \]
  where each set $n$ is $\{0,1,\ldots,n-1\}$ and $\res{\Omega}{n}(k)=\min(n,k)$. In
  fact this is the \emph{subobject classifier} of $\trees$, usually written $\Omega$.

  $\den{\CoNat}$ is the constant object $\Delta(\mathbb{N}+\{\infty\})$.
\end{enumerate}
\end{exa}

\begin{lem}\label{lem:rec_types}
  The interpretation of a recursive type is isomorphic to the
  interpretation of its unfolding:
  $\den{\mu\alpha.A}(\vec{W})\iso \den{A[\mu\alpha.A/\alpha]}(\vec{W})$.
\qed
\end{lem}

\begin{lem}\label{lem:const_types}
Constant types denote constant objects in $\trees$.
\end{lem}
\proof
By induction on type formation, with $\LATER A$ case omitted, $\blacksquare A$ a base
case, and $\mu\alpha.A$ considered only where $\alpha$ is not free in $A$.
\qed

Note that the converse does not apply; for example $\den{\LATER 1}$ is a constant
object.

\begin{defi}\label{def:terms_denote}
We interpret typing contexts $\Gamma=x_1 : A_1, \ldots, x_n :A_n$ in the usual way as
$\trees$-objects $\den{\Gamma}\defeq \den{A_1}\times\cdots\times\den{A_n}$, and
hence interpret typed terms-in-context $\Gamma\vdash t:A$ as $\trees$-arrows
$\den{\Gamma\vdash t:A}:\den{\Gamma}\to\den{A}$ (usually written $\den{t}$) as
follows.

$\den{x}$ is the projection $\den{\Gamma}\times\den{A}\to\den{A}$. $\den{\ZERO}$
and $\den{\SUCC t}$ are as obvious. Term-formers for products and function spaces
are interpreted via the cartesian closed structure of $\trees$, and for sums via its
coproducts. Exponentials are not merely pointwise, so we give the definitions explicitly:
\begin{itemize}
  \item $\den{\lambda x.t}_i(\gamma)_j$ maps $a\mapsto
    \den{\Gamma,x:A\vdash t:B}_j(\restriction_j(\gamma),a)$;
  \item $\den{t_1t_2}_i(\gamma)=(\den{t_1}_i(\gamma)_i) \circ
    \den{t_2}_i(\gamma)$;
\end{itemize}

$\den{\FOLD t}$ and $\den{\UNFOLD t}$ are defined via composition with the 
isomorphisms of Lemma~\ref{lem:rec_types}.
$\den{\NEXT t}$ and $\den{\UNBOX t}$ are defined by composition with the natural
transformations introduced in Definition~\ref{def:functors}. The final cases are
\begin{itemize}
\item
  $\den{t_1\APP t_2}_1$ is defined uniquely at the trivial first stage of the denotation of
  a later type; $\den{t_1\APP t_2}_{i+1}(\gamma)\defeq
  (\den{t_1}_{i+1}(\gamma)_i)\circ\den{t_2}_{i+1}(\gamma)$.
\item
  $\den{\PREV [x_1\explsubst t_1,\ldots] .t}_i(\gamma)\defeq\den{t}_{i+1}(\den{t_1}_i
  (\gamma),\ldots)$, where $\den{t_1}_i(\gamma)\in\den{A_1}_i$ is also in
  $\den{A_1}_{i+1}$ by Lemma~\ref{lem:const_types};
\item
  $\den{\BOX [x_1\explsubst t_1,\ldots].t}_i(\gamma)_j=
  \den{t}_j(\den{t_1}_i(\gamma),
  \ldots)$, again using Lemma~\ref{lem:const_types};
\item
  Let $\den{t}_j(\den{t_1}_i(\gamma),\ldots,\den{t_n}_i(\gamma))$ (which is
  well-defined by Lemma~\ref{lem:const_types}) be $[a_j,d]$ as $j$ ranges, recalling that
  $d\in\{1,2\}$ is the same for all $i$ by naturality. Define $a$ to be the arrow $1\to
  \den{A_d}$ that has $j$'th element $a_j$. Then $\den{\BOXSUM[\vec{x}\explsubst
  \vec{t}\,].t}_i(\gamma)\defeq[a,d]$.
\end{itemize}
\end{defi}

\begin{lem}\label{lem:subst}
Take typed terms in context $x_1:A_1,\ldots,x_m:A_m\vdash t:A$ and $\Gamma
\vdash t_k:A_k$ for all $1\leq k\leq m$. Then
$\den{t[\vec{t}/\vec{x}]}_i(\gamma)=\den{t}_i(\den{t_1}_i(\gamma),\ldots,
\den{t_m}_i(\gamma))$.
\end{lem}
\proof
By induction on the typing of $t$. We present the cases particular to our calculus.

$\NEXT t$: case $i=1$ is trivial.
$\den{\NEXT t[\vec{t}/\vec{x}]}_{i+1}(\gamma)=
\res{\den{A}}{i}\circ\den{t[\vec{t}/\vec{x}]}_{i+1}(\gamma)$ by definition, which is
$\res{\den{A}}{i}\circ\den{t}_{i+1}(\den{t_1}_{i+1}(\gamma),\ldots)$ by induction,
which is $\den{\NEXT t}_{i+1}(\den{t_1}_{i+1}(\gamma),\ldots)$.

$\den{(\PREV[\vec{y}\explsubst\vec{u}].t)[\vec{t}/\vec{x}]}_i(\gamma)=
\den{\PREV[\vec{y}\explsubst\vec{u}[\vec{t}/\vec{x}]].t}_i(\gamma)$, which by
definition is equal to
$\den{t}_{i+1}(\den{u_1[\vec{t}/\vec{x}]}_i(\gamma),\ldots)$, which is
$\den{t}_{i+1}(\den{u_1}_i(\den{t_1}_i(\gamma),\ldots),\ldots)$ by induction, which is
$\den{\PREV[\vec{y}\explsubst\vec{u}].t}_i(\den{t_1}_i(\gamma),\ldots)$.

$u_1\APP u_2$: case $i=1$ is trivial.
$\den{(u_1\APP u_2)[\vec{t}/\vec{x}]}_{i+1}(\gamma)=
(\den{u_1[\vec{t}/\vec{x}]}_{i+1}(\gamma)_i)\circ
\den{u_2[\vec{t}/\vec{x}]}_{i+1}(\gamma)$, which is
$(\den{u_1}_{i+1}(\den{t_1}_{i+1}(\gamma),\ldots)_i)\circ
\den{u_2}_{i+1}(\den{t_1}_{i+1}(\gamma),\ldots)$, which is in turn equal to
$\den{u_1\APP u_2}_{i+1}(\den{t_1}_{i+1}(\gamma),\ldots)$.

$\den{\BOX[\vec{y}\explsubst\vec{u}[\vec{t}/\vec{x}]].t}_i(\gamma)_j=
\den{t}_j(\den{u_1[\vec{t}/\vec{x}]}_i(\gamma),\ldots)$, which is
$\den{t}_j(\den{u_1}_i(\den{t_1}_i(\gamma),\ldots),\ldots)$ by induction, which is
$\den{\BOX[\vec{y}\explsubst\vec{u}].t}_i(\den{t_1}_i(\gamma),\ldots)_j$.

$\den{\UNBOX t[\vec{t}/\vec{x}]}_i(\gamma)=
\den{t[\vec{t}/\vec{x}]}_i(\gamma)_i=
\den{t}_i(\den{t_1}_i(\gamma),\ldots)_i=
\den{\UNBOX t}_i(\den{t_1}_i(\gamma),\ldots)$.

$\BOXSUM[\vec{y}\explsubst\vec{u}{]}.t$: By induction we have
$\den{u_k[\vec{t}/\vec{x}]}_i(\gamma)=\den{u_k}_i(\den{t_1}_i
(\gamma),\ldots)$. Hence $\den{t}_j(\den{u_1[\vec{t}/\vec{x}]}_i(\gamma),\ldots)=
\den{t}_j(\den{u_1}_i(\den{t_1}_i(\gamma),\ldots),\ldots)$ as required.
\qed\smallskip

\begin{thm}[Soundness]\label{lem:soundness}
If $t\redrt u$ then $\den{t}=\den{u}$.
\end{thm}
\proof
We verify the reduction rules of Definition~\ref{def:redrule}; extending this to any
evaluation context, and to $\redrt$, is easy. The product reduction case is standard,
and function case requires Lemma~\ref{lem:subst}. $\UNFOLD\FOLD$ is the application of
mutually inverse arrows.

$\den{\PREV[\vec{x}\explsubst\vec{t}\,].t}_i=\den{t}_{i+1}(\den{t_1}_i,\ldots)$. Each
$t_k$ in the explicit substitution is closed, so is denoted by an arrow from $1$ to a
constant $\trees$-object, so by naturality $\den{t_k}_i=\den{t_k}_{i+1}$. $\den{t}_{i+1}
(\den{t_1}_{i+1},\ldots)=\den{t[\vec{t}/\vec{x}]}_{i+1}$ by Lemma~\ref{lem:subst},
which is $\den{\PREV t[\vec{t}/\vec{x}]}_i$.

$\den{\PREV\NEXT t}_i=\den{\NEXT t}_{i+1}=\den{t}_i$.

With $\APP$-reduction, index $1$ is trivial.
$\den{\NEXT t_1\APP\NEXT t_2}_{i+1}=
(\den{\NEXT t_1}_{i+1})_i\circ\den{\NEXT t_2}_{i+1}=
(\res{\den{A\to B}}{i}\circ \den{t_1}_{i+1})_i\circ
\res{\den{A}}{i}\circ \den{t_2}_{i+1}=
(\den{t_1}_i\circ\res{1}{i})_i\circ\den{t_2}_i\circ \res{1}{i}$ by naturality, which is
$(\den{t_1}_i)_i\circ\den{t_2}_i=
\den{t_1t_2}_i=
\den{t_1t_2}_i\circ\res{1}{i}=
\res{\den{B}}{i}\circ \den{t_1t_2}_{i+1}=
\den{\NEXT(t_1t_2)}_{i+1}$.

$\den{\UNBOX(\BOX[\vec{x}\explsubst\vec{t}\,].t)}_i=(\den{\BOX
  [\vec{x}\explsubst\vec{t}\,].t}_i)_i
=\den{t}_i(\den{t_1}_i,\ldots)=\den{t[\vec{t}/\vec{x}]}_i$.

$\BOXSUM$-reduction:
Because each $\den{A_k}$ is a constant object (Lemma~\ref{lem:const_types}),
$\den{t_k}_i=\den{t_k}_j$ for all $i,j$. Hence $\den{\BOXSUM[\vec{x}\explsubst
\vec{t}\,].t}_i$ is defined via components $\den{t}_j(\den{t_1}_j,\ldots)$ and
$\den{\BOXSUM t[\vec{t}/\vec{x}]}$ is defined via components $\den{t[\vec{t}/
\vec{x}]}_j$. These are equal by Lemma~\ref{lem:subst}.
$\den{\BOXSUM\IN_d t}_i$ is the $d$'th injection into the function with $j$'th
component $\den{t}_j$, and likewise for $\den{\IN_d\BOX t}_i$.
\qed

\subsection{Adequacy and Normalisation}

We now define a logical relation between our denotational semantics and
terms, from which both normalisation and adequacy will follow. Doing this inductively
proves rather delicate, because induction on size will not support reasoning
about our values, as $\FOLD$ refers to a larger type in its premise.
This motivates a notion
of \emph{unguarded size} under which $A[\mu\alpha.A/\alpha]$ is `smaller' than $\mu
\alpha.A$. But under this metric $\LATER A$ is smaller than $A$, so $\NEXT$ now
poses a problem. But the meaning of $\LATER A$ at index $i+1$ is determined by
$A$ at index $i$, and so, as in Birkedal et al.~\cite{Birkedal:Metric}, our relation will
also induct on
index. This in turn creates problems with $\BOX$, whose meaning refers to all indexes
simultaneously, motivating a notion of \emph{box depth}, allowing us finally to
attain well-defined induction.

\begin{defi}
The \emph{unguarded size} $\usize$ of an open type follows the obvious definition for
type size, except that $\usize(\LATER A)=0$.

The \emph{box depth} $\boxd$ of an open type is
\begin{itemize}
  \item $\boxd(A)=0$ for $A\in\{\alpha,\EMPTY,\ONE,\NAT\}$;
  \item $\boxd(A\times B)=\min(\boxd(A),\boxd(B))$, and similarly for
    $A+B,A\to B$;
  \item $\boxd(\mu\alpha.A)=\boxd(A)$, and similarly for $\boxd(\LATER A)$;
  \item $\boxd(\blacksquare A)=\boxd(A)+1$.
\end{itemize}
\end{defi}

\begin{lem}\label{lem:boxdepth}\hfill
\begin{enumerate}
\item
  $\alpha$ guarded in $A$ implies $\usize(A[B/\alpha])\leq \usize(A)$.
  \label{lem:boxdepth-i}
\item
  $\boxd(B)\leq \boxd(A)$ implies $\boxd(A[B/\alpha])\leq \boxd(A)$
  \label{lem:boxdepth-ii}
\end{enumerate}
\end{lem}
\proof
By induction on the construction of the type $A$.

$(i)$ follows with only interesting case the variable case -- $A$ cannot be $\alpha$
because of the requirement that $\alpha$ be guarded in $A$.

$(ii)$ follows with interesting cases: variable case enforces $bd(B)=0$; binary
type-formers $\times,\to$ have for example $\boxd(A_1)\geq \boxd(A_1\times A_2)$,
so $\boxd(A_1)\geq bd(B)$ and the induction follows; $\blacksquare A$ by construction
has no free variables.
\qed

\begin{defi}\label{Def:log_rel}
The family of relations $\lrel{i}{A}$, indexed by closed types $A$ and positive integers
$i$, relates elements of the semantics $a\in\den{A}_i$ and closed typed terms $t:A$
and is defined as
\begin{itemize}
\item $n \lrel{i}{\NAT} t$ iff $t \redrt \SUCC^n\ZERO$;
\item $\ast \lrel{i}{\ONE} t$ iff $t\redrt\UNIT$;
\item $(a_1,a_2)\lrel{i}{A_1\times A_2} t$ iff $t\redrt\langle t_1,t_2\rangle$ and
  $a_1\lrel{i}{A_1}t_1$ and $a_2\lrel{i}{A_2}t_2$;
\item $[a,d]\lrel{i}{A_1+A_2} t$ iff $t\redrt \IN_d u$ for $d\in\{1,2\}$, and $a
  \lrel{i}{A_d}u$.
\item $f\lrel{i}{A\to B} t$ iff $t\redrt\lambda x.s$ and for all $j\leq i$, $a\lrel{j}{A}u$
  implies $f_j(a)\lrel{j}{B} s[u/x]$;
\item $a\lrel{i}{\mu\alpha.A}t$ iff $t\redrt \FOLD u$ and $h_i(a)\lrel{i}{A[\mu
  \alpha.A/\alpha]}u$, where $h$ is the ``unfold'' isomorphism for the recursive type
  (ref. Lemma~\ref{lem:rec_types});
\item $a\lrel{i}{\LATER A} t$ iff $t\redrt\NEXT u$ and, where $i>1$, $a\lrel{i-1}{A}u$.
\item $a\lrel{i}{\blacksquare A} t$ iff $t\redrt\BOX u$ and
  for all $j$, $a_j\lrel{j}{A}u$;
\end{itemize}
Note that $\lrel{i}{\EMPTY}$ is (necessarily) everywhere empty.

The above is well-defined by induction on the lexicographic ordering on box depth, then
index, then unguarded size. First, the $\blacksquare$ case strictly decreases box depth,
and no other case increases it (ref. Lemma~\ref{lem:boxdepth}.\ref{lem:boxdepth-ii} for
$\mu$-types). Second, the $\LATER$ case strictly decreases index, and no other case
increases it (disregarding $\blacksquare$). Finally, all other cases strictly decrease
unguarded size, as seen via Lemma~\ref{lem:boxdepth}.\ref{lem:boxdepth-i} for
$\mu$-types.
\end{defi}\newpage

\begin{lem}\label{lem:lrel_and_red}
If $t\redrt u$ and $a\lrel{i}{A}u$ then $a\lrel{i}{A}t$.
\end{lem}
\proof
All cases follow similarly; consider $A_1\times A_2$. $(a_1,a_2)
\lrel{i}{A_1\times A_2}u$ implies $u\redrt\langle t_1,t_2\rangle$, where this value
obeys some property. But then $t\redrt\langle t_1,t_2\rangle$ similarly.
\qed

\begin{lem}\label{lem:res_and_rel}
$a\lrel{i+1}{A}t$ implies $\res{\den{A}}{i}(a)\lrel{i}{A}t$.
\end{lem}
\proof
Cases $\NAT,\ONE,\EMPTY$ are trivial. Cases $\times$ and $+$ follow by induction
because restrictions are defined pointwise. Case $\mu$ follows by induction and the
naturality of the isomorphism $h$. Case $\blacksquare A$ follows because
$\res{\den{\blacksquare A}}{i}(a)=a$.

For $A\to B$ take $j\leq i$ and $a'\lrel{j}{A}u$. By the downwards closure in the
definition of $\lrel{i+1}{A\to B}$ we have $f_j(a')\lrel{j}{B}s[u/x]$. But
$f_j=(\res{\den{A\to B}}{i}(f))_j$.

With $\LATER A$, case $i=1$ is trivial, so take $i=j+1$. $a\lrel{j+2}{\LATER A}t$
means $t\redrt\NEXT u$ and $a\lrel{j+1}{A}u$, so by induction $\res{\den{A}}{j}(a)
\lrel{j}{A}u$, so $\res{\den{\LATER A}}{j+1}(a)\lrel{j}{A}u$ as required.
\qed

\begin{lem}\label{lem:constrel}
If $a\lrel{i}{A}t$ and $A$ is constant, then $a\lrel{j}{A}t$ for all $j$.
\end{lem}
\proof
Easy induction on types, ignoring $\LATER A$ and treating $\blacksquare A$ as a base
case.
\qed

We may now turn to the proof of the Fundamental Lemma.

\begin{lem}[Fundamental Lemma]\label{lem:ad}
Take $\Gamma=(x_1:A_1,\ldots,x_m:A_m)$, $\Gamma\vdash t:A$, and closed typed
terms $t_k:A_k$ for $1\leq k\leq m$. Then for all $i$, if $a_k\lrel{i}{A_k}t_k$ for all $k$,
then
\[
  \den{\Gamma\vdash t:A}_i(\vec{a})\,\lrel{i}{A}\,t[\vec{t}/\vec{x}].
\]
\end{lem}
\proof
By induction on the typing $\Gamma\vdash t:A$. $\UNIT,\ZERO$ cases are trivial, and
$\langle u_1,u_2\rangle,\IN_d t,\FOLD t$ cases follow by easy induction.

$\SUCC t$: If $t[\vec{t}/\vec{x}]$ reduces to $\SUCC^l\ZERO$ for some $l$ then
$\SUCC t[\vec{t}/\vec{x}]$ reduces to $\SUCC^{l+1}\ZERO$, as we may reduce
under the $\SUCC$.

$\pi_d t$ for $d\in\{1,2\}$: If $\den{t}_i(\vec{a})\lrel{i}{A_1\times A_2}t
[\vec{t}/\vec{x}]$ then $t[\vec{t}/\vec{x}]\redrt\langle u_1,u_2\rangle$ and $u_d$ is
related to the $d$'th projection of $\den{t}_i(\vec{a})$. But then $\pi_d t
[\vec{t}/\vec{x}]\redrt\pi_d\langle u_1,u_2\rangle\red u_d$, so
Lemma~\ref{lem:lrel_and_red} completes the case.

$\ABORT$: The induction hypothesis states that $\den{t}_k(\vec{a})\lrel{k}{\EMPTY}
t[\vec{t}/\vec{x}\,]$, but this is not possible, so the statement holds vacuously.

$\CASE t\OF y_1.u_1;y_2.u_2$: If $\den{t}_i(\vec{a})\lrel{i}{A_1+A_2}t[\vec{t}/
\vec{x}]$ then $t[\vec{t}/\vec{x}\,]\redrt\IN_d u$ for some $d\in\{1,2\}$, with
$\den{t}_i(\vec{a})=[a,d]$ and $a\lrel{i}{A_d}u$. Then $\den{u_d}_i(\vec{a},a)
\lrel{k}{A}u_d[\vec{t}/\vec{x},u/y_d]$. Now we have that $(\CASE t\OF y_1.u_1;y_2.u_2)[\vec{t}/
\vec{x}]\redrt\CASE\IN_d u\OF y_1.(u_1[\vec{t}/\vec{x}]);y_2.(u_2[\vec{t}/\vec{x}])$,
which in turn reduces to $u_d[\vec{t}/\vec{x},u/y_i]$, and
Lemma~\ref{lem:lrel_and_red} completes.

$\lambda x.t$: Taking $j\leq i$ and $a\lrel{j}{A}u$, we must show that
$\den{\lambda x.t}_i(\vec{a})_j(a)\lrel{j}{B}t[\vec{t}/\vec{x}][u/x]$. The left hand
side is $\den{t}_j(\restriction_j(\vec{a}),a)$. For each $k$, $a_k
\hspace{-0.3em}\restriction_j\hspace{-0.3em}\lrel{j}{A_k}t_k$ by
Lemma~\ref{lem:res_and_rel}, and induction completes the case.

$u_1u_2$: By induction $u_1[\vec{t}/\vec{x}]\redrt \lambda x.s$ and
$\den{u_1}_k(\vec{a})_k(\den{u_2}_k(\vec{a}))\lrel{i}{B}s[u_2[\vec{t}/\vec{x}]/x]$.
Now we have $(u_1u_2)\redrt(\lambda x.s)(u_2[\vec{t}/\vec{x}])\red
s[u_2[\vec{t}/\vec{x}]/x]$, and Lemma~\ref{lem:lrel_and_red} completes.

$\UNFOLD t$: we reduce under $\UNFOLD$,
then reduce $\UNFOLD\FOLD$, then use Lemma~\ref{lem:lrel_and_red}.

$\NEXT t$: Trivial for index $1$. For $i=j+1$, if each $a_k\lrel{j+1}{A_k}t_k$ then by
Lemma~\ref{lem:res_and_rel} $\res{\den{A_k}}{j}(a_k)\lrel{j}{A_k}
t_k$. Then by induction $\den{t}_j\circ\res{\den{\Gamma}}{j}(\vec{a})\lrel{j}{A}t[\vec{t}/\vec{x}]$, whose left side is by naturality
$\res{\den{A}}{j}\circ\den{t}_{j+1}(\vec{a})=\den{\NEXT t}_{j+1}(\vec{a})$.

$\PREV[\vec{y}\explsubst\vec{u}].t$: $\den{u_k}_i(\vec{a})\lrel{i}{A_k}u_k[\vec{t}/
\vec{x}]$ by induction, so $\den{u_k}_i(\vec{a})\lrel{i+1}{A_k}u_k[\vec{t}/
\vec{x}]$ by Lemma~\ref{lem:constrel}. Then $\den{t}_{i+1}(\den{u_1}_i(\vec{a}),
\ldots)\lrel{i+1}{\LATER A} t[u_1[\vec{t}/\vec{x}]/y_1,\ldots]$ by induction, so we
have $t[u_1[\vec{t}/\vec{x}]/y_1,\ldots]\redrt\NEXT s$ with $\den{t}_{i+1}
(\den{u_1}_k(\vec{a}),\ldots)\lrel{i}{A}s$. The left hand side is $\den{\PREV[\vec{y}\explsubst\vec{u}] 
.t}_i(\vec{a})$, while
$\PREV[\vec{y}\explsubst\vec{u}[\vec{t}/\vec{x}]].t\red \PREV
t[u_1[\vec{t}/\vec{x}]/y_1,\ldots]
\redrt\PREV\NEXT s\red s$, so Lemma~\ref{lem:lrel_and_red} completes.

$u_1\APP u_2$: Index $1$ is trivial so set $i=j+1$. $\den{u_2}_{j+1}(\vec{a})
\lrel{j+1}{\LATER A}u_2[\vec{t}/\vec{x}]$ implies $u_2[\vec{t}/\vec{x}]\redrt\NEXT s_2$ with
$\den{u_2}_{j+1}(\vec{a})\lrel{j}{A}s_2$. Similarly $u_1\redrt\NEXT s_1$ and
$s_1\redrt \lambda x.s$ with
$(\den{u_1}_{j+1}(\vec{a})_j)\circ\den{u_2}_{j+1}(\vec{a})\lrel{j}{B}s[s_2/x]$.
The left hand side is exactly $\den{u_1\APP u_2}_{j+1}(\vec{a})$. Now $u_1\APP u_2
\redrt \NEXT s_1\APP u_2\redrt \NEXT s_1\APP \NEXT s_2\red \NEXT(s_1s_2)$, and
$s_1s_2\redrt (\lambda x.s)s_2\red s[s_2/x]$, completing the proof.

$\BOX[\vec{y}\explsubst\vec{u}].t$: To show $\den{\BOX[\vec{y}\explsubst\vec{u}].t}_i
(\vec{a})\lrel{i}{\blacksquare A}\BOX[\vec{y}\explsubst\vec{u}].t)[\vec{t}/\vec{x}]$, we
observe that the right hand side reduces in one step to $\BOX t[u_1[\vec{t}/
\vec{x}]/y_1,\ldots]$. The $j$'th element of the left hand side is $\den{t}_j
(\den{u_1}_k(\vec{a}),\ldots)$. We need to show this is related by $\lrel{j}{A}$ to
$ t[u_1[\vec{t}/\vec{x}]/y_1,\ldots]$; this follows by Lemma~\ref{lem:constrel} and
induction.

$\UNBOX t$: By induction $t[\vec{t}/\vec{x}]\redrt\BOX u$, so $\UNBOX t[\vec{t}/
\vec{x}]\redrt\UNBOX\BOX u\red u$. By induction $\den{t}_i(\vec{a})_i\lrel{i}{A}u$,
so $\den{\UNBOX t}_i(\vec{a})\lrel{i}{A}u$, and Lemma~\ref{lem:lrel_and_red}
completes.

$\BOXSUM[\vec{y}\explsubst\vec{u}].t$: $\den{u_k}_i(\vec{a})\lrel{i}{A_k}u_k[\vec{t}/
\vec{x}]$ by induction, so $\den{u_k}_i(\vec{a})\lrel{j}{A_k}u_k[\vec{t}/\vec{x}]$
for any $j$ by Lemma~\ref{lem:constrel}. By induction $\den{t}_j(\den{u_1}_k
(\vec{a}),\ldots)\lrel{j}{B_1+B_2}t[u_1[\vec{t}/\vec{x}\,]/y_1,\ldots]$. If
$\den{t}_j(\den{u_1}_k(\vec{a}),\ldots)$ is some $[b_j,d]$ we have $t[u_1[\vec{t}/
\vec{x}]/y_1,\ldots]\redrt\IN_d s$ with $b_j\lrel{j}{B_d}s$. Now $(\BOXSUM[\vec{y}
\explsubst\vec{u}].t)[\vec{t}/\vec{x}\,]\red \BOXSUM t[u_1[\vec{t}/\vec{x}\,]/y_1,\ldots]
\redrt \BOXSUM\IN_d s$, which finally reduces to $\IN_d\BOX s$, which yields the
result.
\qed

\begin{thm}[Adequacy and Normalisation]
  \label{thm:ad_norm}\hfill
\begin{enumerate}
  \item\label{it:ad_norm1} For all closed terms $\vdash t:A$ it holds that $\den{t}_i\lrel{i}{A} t$;
  \item\label{it:ad_norm2} $\den{\vdash t:\NAT}_i=n$ implies $t\redrt \SUCC^n\ZERO$;
    \label{thm:ad_norm:den-implies-red}
  \item\label{it:ad_norm3} All closed typed terms evaluate to a value.
\end{enumerate}
\end{thm}
\proof
\eqref{it:ad_norm1} specialises Lemma~\ref{lem:ad} to closed
types. \eqref{it:ad_norm2} and \eqref{it:ad_norm3} hold by \eqref{it:ad_norm1} and
inspection of Definition~\ref{Def:log_rel}.
\qed

\begin{defi}
Typed \emph{contexts} with typed holes are defined as obvious. Two terms
$\Gamma \vdash t : A, \Gamma \vdash u : A$
are \emph{contextually equivalent}, written $t\ceq u$, if for all
well-typed \emph{closing} contexts $C$ of type $\NAT$, the terms $C[t]$ and $C[u]$ reduce to the same value.
\end{defi}

\begin{cor}
  \label{cor:adequacy:den-implies-ctxeq}
  $\den{t}=\den{u}$ implies $t\ceq u$.
\end{cor}
\proof
$\den{C[t]}=\den{C[u]}$ by compositionality of the denotational semantics. Then
by Theorem~\ref{thm:ad_norm}.\ref{thm:ad_norm:den-implies-red} they reduce to the
same value.
\qed


\section{Logic for the Guarded Lambda Calculus}
\label{sec:logic}

In this section we will discuss the internal logic of the topos of trees, show that it yields
a program logic $\logiclambdanext$ which supports reasoning about the contextual
equivalence of $\glambda$-programs, remark on some properties of this program logic,
and give some example proofs.

\subsection{From Internal Logic to Program Logic}

$\trees$ is a presheaf category, and so a topos, and so its internal logic provides a
model of higher-order logic with equality~\cite{Maclane:Sheaves}. The internal logic of
$\trees$ has been explored
elsewhere~\cite{Birkedal-et-al:topos-of-trees,Litak:Constructive,Clouston:Sequent},
but to motivate the results of this section we make some observations here.

As discussed in Example~\ref{ex:denote_streams}.\ref{ex:gCoNat}, the subobject
classifier $\Omega$ is exactly the denotation of the \emph{guarded conatural numbers}
$\gCoNat$, as defined in the $\glambda$-calculus in Section~\ref{subsec:boxplus}.
The propositional connectives can then be defined via $\glambda$-functions
on the guarded conaturals: false $\bot$ is $\mathsf{cozero}$, as defined in
Section~\ref{subsec:boxplus}; true $\top$ is $\mathsf{infinity}$; conjunction $\land$
is a minimum function readily definable on pairs of guarded conaturals; $\lnot$ is
\[
  \lambda n.\CASE(\UNFOLD n)\OF x_1.\mathsf{infinity};\,
    x_2.\mathsf{cozero} \;:\; \gCoNat\to\gCoNat
\]
and so on. The
connectives $\forall x : A$, $\exists x : A$, and $=_A$ cannot be expressed as
$\glambda$-functions for an arbitrary $\glambda$-type $A$, but are definable as (parametrised)
operations on $\Omega$ in the usual way~\cite[Section IV.9]{Maclane:Sheaves}.

Along with the standard connectives we can define a \emph{modality}
$\later$, whose action on the subobject classifier corresponds precisely to the function
$\mathsf{cosucc}$ on guarded conaturals defined in Section~\ref{subsec:boxplus}. We
call this modality `later', overloading our name for our type-former $\LATER$, and the
functor on $\trees$ with the same name and symbol introduced in
Definition~\ref{def:functors}.\ref{def:functorslater}. This overloading is justified
by a tight relationship between these concepts which we will investigate below. For now,
note that $\mathsf{cosucc}$ can be defined
as a composition of functions $\lift\comp \NEXT$, where $\lift$%
\footnote{called $\mathsf{succ}$ by Birkedal et
al.~\cite{Birkedal-et-al:topos-of-trees}; we avoid this because of the clash with
the name for a term-former.}
is a function $\LATER\Omega\to\Omega$ definable in the $\glambda$ calculus as
\[
  \lambda n.\FOLD(\IN_2 n) \;:\; \LATER\gCoNat\to\gCoNat
\]
Further, $\mathsf{infinity}$ is $\Theta\lift$. We will make use of this $\lift$ function later
in this section.

Returning to the propositional connectives, double negation $\lnot\lnot$ corresponds to
the $\glambda$-function
\[
 \lambda n.\CASE(\UNFOLD n)\OF x_1.\mathsf{cozero};\,
    x_2.\mathsf{infinity} \;:\; \gCoNat\to\gCoNat
\]
Now consider the poset $\Sub{X}$ of subobjects of $X$, which are pointwise subsets
whose restriction maps are determined by the restriction maps of $X$; or equivalently,
characteristic arrows $X\to\Omega$. The function $\lnot\lnot:\Omega\to\Omega$
extends to a monotone function $\Sub{X}\to\Sub{X}$ by composition with
characteristic arrows as obvious. This function preserves joins,
and so by the adjoint functor theorem for posets has a right adjoint
$\Sub{X}\to\Sub{X}$, which we write $\always$ and call
`always'~\cite{Bizjak-et-al:countable-nondet-internal}. The notational
similarity with the type-former and functor $\blacksquare$ is, as with $\later$ and
$\LATER$, deliberate and will be explored further. First, we can offer a more concrete
definition of $\always$:

\begin{defi}\label{defi:always}\hfill
\begin{itemize}
\item
Take a $\trees$-object $X$, positive integer $m$, and element $x\in X_m$, and recall
that for any $n\geq m$ the function $\restriction_m:X_n\to X_m$ is defined by
composing restriction functions. Then the \emph{height} of $x$ in $X$, written
$\height{x}{X}$, is the largest integer $n\geq m$ such that there exists $y\in
X_n$ with $\restriction_m(y)=x$, or $\infty$ if there is no such largest $n$.
\item
Given a subobject $Y$ of $X$, the characteristic arrow of the subobject
$\always Y$ of $X$ is defined as
\[
  (\chi_{\always Y})_n(x) \;=\; \begin{cases}
    (\chi_Y)_n(x) & \height{x}{Y}=\height{x}{X} \\
    0 & \mbox{otherwise.}
  \end{cases}
\]
\end{itemize}
\end{defi}\medskip

\noindent The condition regarding the height of elements allows the modality $\always$ to
reflect the global, rather than pointwise, structure of a subobject. For example,
considering the object $\LATER 0$, which is a singleton at its first stage
and empty set at all later stages, as a subobject of the terminal object $1$, the
subobject $\always(\LATER 0)$ is $0$.

\begin{exa}
A proposition $\phi$ with no free variables corresponds in the internal logic of $\trees$
to an arrow $1\to\Omega$, which as we have seen in turn corresponds to a guarded
conatural number. The proposition $\always\phi$ also corresponds a guarded conatural
number, so we can see the action of $\always$ \emph{on closed propositions} as
arising from a function $\mathbb{N}+\{\infty\}\to\mathbb{N}+\{\infty\}$ defined by
\begin{equation}\label{eq:always_on_conats}
  \always(n) \;=\; \begin{cases}
    \infty & n=\infty \\
    0 & \mbox{otherwise.}
  \end{cases}
\end{equation}
This is a perfectly good function in $\sets$, but it does \emph{not} correspond to an
$\trees$-arrow $\Omega\to\Omega$, because it is hopelessly unproductive -- we need
to make infinitely many observations of the input before we decide anything about the
output. Similarly, we cannot define a function of the type $\gCoNat\to\gCoNat$ in the
$\glambda$-calculus with this behaviour.

The case where we have a subobject $Y$ of a \emph{constant} object $X$ is
similar to the case of subobjects of $1$ -- the characteristic function of $\always Y$
maps each element $x$ of $X$ to a conatural number, which is
then composed with the $\always$ function \eqref{eq:always_on_conats}.
\end{exa}

Note further than $\always$ does not commute with substitution; in particular, given a
substitution $\sigma$, $\always(\phi\sigma)$ does not necessarily imply
$(\always\phi)\sigma$. However these formulae \emph{are} equivalent if $\sigma$ is a
substitution between constant contexts. In practice we will use $\always$ only in
constant context.

We may now proceed to the definition of the program logic $\logiclambdanext$:

\begin{defi}
$\logiclambdanext$ is the typed higher order logic with equality defined by the internal
logic of $\trees$, whose types and function symbols are the types and term-formers of
the $\glambda$-calculus, interpreted in $\trees$ as in Section~\ref{sec:sound_denot},
and further extended by the modalities $\later,\always$.

We write $\Gamma \mid \Xi \vdash \phi$ where the proposition $\phi$ with term
variables drawn from the context $\Gamma$ is entailed by the set of propositions
$\Xi$. Note that we use the symbol $\Omega$ for the type of propositions,
although this is precisely the denotation of the guarded conatural numbers.
\end{defi}

This logic may be used to prove contextual equivalence of programs:

\begin{thm}
  \label{thm:prop-equality-ctx-equiv}
  Let $t_1$ and $t_2$ be two $\lambdanext$ terms of type $A$ in context $\Gamma$. If
  the sequent $\Gamma \mid \emptyset \vdash t_1 =_{A} t_2$
  is provable, then $t_1$ and $t_2$ are contextually equivalent.
\end{thm}
\proof
  Recall that equality in the internal logic of a topos is just equality of
  morphisms. Hence $t_1$ and $t_2$ denote same morphism from $\den{\Gamma}$ to
  $\den{A}$. Adequacy (Corollary~\ref{cor:adequacy:den-implies-ctxeq}) then implies
  that $t_1$ and $t_2$ are contextually equivalent.
\qed

\subsection{Properties of the Logic}\label{sec:logic_props}

The definition of the logic $\logiclambdanext$ from the previous section establishes its
syntax, and semantics in the topos of trees, without giving much sense of how proofs
might be constructed. Clouston and Gor\'e~\cite{Clouston:Sequent} have
provided a sound and complete sequent calculus, and hence decision procedure, for the
fragment of the internal logic of $\trees$ with propositional connectives and $\later$,
but the full logic $\logiclambdanext$ is considerably more expressive than this; for
example it is not decidable~\cite{Mogelberg:reduction}. In this section we will establish
some reasoning principles for $\logiclambdanext$, which will assist us in the next
section in constructing proofs about $\glambda$-programs.

We start by noting that the usual $\beta\eta$-laws and commuting conversions for
the $\lambda$-calculus with products, sums, and iso-recursive types hold. These may
be extended with new equations for the new $\lambdanext$-constructs, sound in the
model $\trees$, as listed in Figure~\ref{fig:additional-equations}.

\begin{figure}
  \centering
  \begin{mathpar}
     \inferrule*{%
      \vec{x}:\vec{A} \vdash t: A \\
      \Gamma\vdash \vec{t}:\vec{A} }{%
      \Gamma \vdash \PREV [x_1\explsubst t_1,\ldots,x_n\explsubst t_n].(\NEXT t)=t\left[\vec{t}/\vec{x}\right]}
    \and
     \inferrule*{%
      \vec{x}:\vec{A} \vdash t: \LATER A \\
      \Gamma\vdash \vec{t}:\vec{A} }{%
      \Gamma \vdash \NEXT(\PREV [x_1\explsubst t_1,\ldots,x_n\explsubst t_n].t)=t\left[\vec{t}/\vec{x}\right]}
    \and
    \inferrule*{\Gamma \vdash t_1 : A \to B\\
      \Gamma \vdash t_2 : A}{\Gamma \vdash \NEXT t_1 \APP \NEXT t_2 = \NEXT (t_1\, t_2)}
    \and
    \inferrule*{\Gamma \vdash f : \LATER(B \to C) \\
      \Gamma \vdash g : \LATER(A \to B) \\
      \Gamma \vdash t:\LATER A}
      {\Gamma \vdash f\APP(g\APP t) =
      (\NEXT\mathsf{comp})\APP f\APP g\APP t}
    \and
     \inferrule*{%
      \vec{x}:\vec{A} \vdash t: A \\
      \Gamma\vdash \vec{t}:\vec{A} }{%
      \Gamma \vdash \UNBOX(\BOX[\vec{x}\explsubst\vec{t}].t) = t\left[\vec{t}/{\vec{x}}\right]}
    \and
     \inferrule*{%
      \vec{x}:\vec{A} \vdash t: \blacksquare A \\
      \Gamma\vdash \vec{t}:\vec{A} }{%
      \Gamma \vdash \BOX[\vec{x}\explsubst\vec{t}].\UNBOX t = t\left[\vec{t}/{\vec{x}}\right]}
    \and
     \inferrule*{%
      \vec{x}:\vec{A} \vdash t: A \\
      \Gamma\vdash \vec{t}:\vec{A} }{%
      \Gamma \vdash \BOXSUM[\vec{x}\explsubst\vec{t}].\IN_1 t = \IN_1\BOX[\vec{x}\explsubst\vec{t}]. t}
    \and
     \inferrule*{%
      \vec{x}:\vec{A} \vdash t: B \\
      \Gamma\vdash \vec{t}:\vec{A} }{%
      \Gamma \vdash \BOXSUM[\vec{x}\explsubst\vec{t}].\IN_2 t = \IN_2\BOX[\vec{x}\explsubst\vec{t}]. t}
    \and
    \mprset{flushleft}
     \inferrule*{%
      \vec{x}:\vec{A} \vdash t: A+B \\
      \Gamma\vdash \vec{t}:\vec{A} \\
      \Gamma,z_1:A\vdash u_1:C \\
      \Gamma,z_2:B\vdash u_2:C }{%
      \Gamma \vdash  \CASE(\BOXSUM[\vec{x}\explsubst\vec{t}].t)\OF y_1.u_1[\UNBOX y_1/z_1]; y_2.u_2[\UNBOX y_2/z_2] \\
      \hphantom{\Gamma \vdash} = \CASE(t[\vec{t}/\vec{x}])\OF z_1.u_1;z_2.u_2}
    \and
     \inferrule*{%
      \vec{x}:\vec{A} \vdash t: C+D \\
      \Gamma\vdash \vec{t}:\vec{A} \\
      \vec{x}:\vec{A},y_1:C \vdash u_1: A+B \\
      \vec{x}:\vec{A},y_2:D \vdash u_2: A+B }{%
      \Gamma \vdash  \BOXSUM[\vec{x}\explsubst\vec{t}].\CASE t\OF y_1.u_1;y_2.u_2 \\
      \hphantom{\Gamma \vdash} =
      \CASE(t[\vec{t}/\vec{x}])\OF y_1.\BOXSUM[\vec{x},y_1\explsubst\vec{t},y_1].u_1;
      y_2.\BOXSUM[\vec{x},y_2\explsubst\vec{t},y_2].u_2}
    \and    
     \inferrule*{%
      \vec{x}:\vec{A} \vdash t: \LATER A \\
      \vec{y}:\vec{B}\vdash \vec{t}:\vec{A} \\
      \Gamma\vdash \vec{u}:\vec{B} }{%
      \Gamma \vdash \PREV[\vec{y}\explsubst\vec{u}].(t[\vec{t}/\vec{x}]) =  \PREV[\vec{x}\explsubst(\vec{t}[\vec{u}/\vec{x}])].t }
    \and
     \inferrule*{%
      \vec{x}:\vec{A} \vdash t: A \\
      \vec{y}:\vec{B}\vdash \vec{t}:\vec{A} \\
      \Gamma\vdash \vec{u}:\vec{B} }{%
      \Gamma \vdash \BOX[\vec{y}\explsubst\vec{u}].(t[\vec{t}/\vec{x}]) =  \BOX[\vec{x}\explsubst(\vec{t}[\vec{u}/\vec{x}])].t }
    \and
     \inferrule*{%
      \vec{x}:\vec{A} \vdash t: A+B \\
      \vec{y}:\vec{B}\vdash \vec{t}:\vec{A} \\
      \Gamma\vdash \vec{u}:\vec{B} }{%
      \Gamma \vdash \BOXSUM[\vec{y}\explsubst\vec{u}].(t[\vec{t}/\vec{x}]) =  \BOXSUM[\vec{x}\explsubst(\vec{t}[\vec{u}/\vec{x}])].t }
  \end{mathpar}
  \caption{Equations between $\lambdanext$-terms in $\logiclambdanext$. Types in
    $\Vec{A},\Vec{B},C,D$ are assumed constant. $\mathsf{comp}$ is the composition
    $\lambda x.\lambda y.\lambda z.x(yz):(B\to C)\to(A\to B)\to(A\to C)$.}
\label{fig:additional-equations}
\end{figure}

Many of the rules of Figure~\ref{fig:additional-equations} are unsurprising, adding
$\eta$-rules to the $\beta$-rules of Definition~\ref{def:redrule}, noting only that in the
case of $\LATER$ we use the rule of equation~\eqref{eq:general_prev_rule}, because
we are here allowing the consideration of open terms. The reduction rule for $\APP$
is joined by the `composition' equality for applicative
functors~\cite{McBride:Applicative}. In addition to the $\beta$-rule
for $\BOXSUM$ of Definition~\ref{defi:boxsum}, which govern how this connective
commutes with the constructors $\IN_1$, $\IN_2$ and $\BOX$, we also add a rule
showing how it interacts with the eliminators $\CASE$ and $\UNBOX$. The next rule
resembles a traditional commuting conversion for $\CASE$ with $\BOXSUM$, but
specialised to hold where the sum $C+D$ on which the case split occurs has constant
type.

There are finally three rules showing how
substitutions can be moved in and out of the explicit substitutions attached to the
term-formers $\PREV$, $\BOX$, and $\BOXSUM$, provided everything is suitably
constant. Because of these operators' binding structure, substituted terms can get
`stuck' inside explicit substitutions and so cannot interact with the terms the operators
are applied to. This is essential for soundness in general, but not where everything is
suitably constant, in which case these rules become essential to further simplifying 
terms. As an example, the rather complicated commuting conversion for Intuitionistic S4
defined by Bierman and de Paiva~\cite{Bierman:Intuitionistic}
\[
  \begin{array}{lcl}
  \BOX[\vec{x}\explsubst\vec{t},\vec{y}\explsubst\vec{u}].(t[\BOX\iota.u/x]) &\approx&
  \BOX[\vec{x}\explsubst\vec{t},x\explsubst (\BOX[\vec{y}\explsubst\vec{u}].u)].t    
  \end{array}
\]
comes as a corollary.

We now pick out a distinguished class of $\trees$-objects and $\glambda$-types that
enjoy extra properties that are useful in some $\logiclambdanext$ proofs.

\begin{defi}
An $\trees$-object is \emph{total and inhabited} if all its restriction functions are
surjective, and all its sets are non-empty.

A $\glambda$-type is \emph{total and inhabited} if its denotation in $\trees$ is total
and inhabited.
\end{defi}\medskip

\noindent In fact we can express this property directly in the internal logic:

\begin{lem}
A type $A$ is total and inhabited iff the formula
\[
  \total{A}\;\defeq\;\forall a' : \LATER  A, \exists a : A, a' =_{\LATER A} \NEXT a
\]
is valid.
\end{lem}
\proof
  The formula $\total{A}$ expresses the \emph{internal surjectivity} of the
  $\trees$-arrow $\NEXT:\den{A}\to\LATER\den{A}$. In any presheaf topos, this holds
  of an arrow precisely when its components are all surjective. It hence suffices to
  show that any $\trees$-object $X$ is total and inhabited iff all the functions of
  $\NEXT:X\to\LATER X$ are surjective: $X_1$ is non-empty iff $!:X_1\to (\LATER
  X)_1=\{\ast\}$ is surjective, all other arrows of $\NEXT$ are the restriction functions
  themselves, and if $X_1$ is non-empty and all restriction functions are surjective, then
  all $X_i$ are non-empty.
\qed

In fact almost all $\glambda$-types are total and inhabited, as the next lemma and
its corollary show:

\begin{lem}
  \label{prop:functor-restr-fp-total}
  Let $F : (\trees^{op}\times\trees)^{n+1}\to\trees$ be a \emph{locally
  contractive}~\cite[Definition II.10]{Birkedal-et-al:topos-of-trees} functor that maps
  tuples of total and inhabited objects to total and inhabited objects, i.e. $F$ restricts to
  the full subcategory $ti\Sl$ of total and inhabited $\trees$-objects.

  Then its fixed point $\mathsf{Fix}(F):(\trees^{op}\times\trees)^{n}\to\trees$ is also
  total and inhabited.
\end{lem}
\proof
  $ti\Sl$ is equivalent to the category of bisected complete non-empty ultrametric spaces
  $\Ml$~\cite[Section 5]{Birkedal-et-al:topos-of-trees}. $\Ml$ is
  known to be an $M$-category in the sense of Birkedal et al.~\cite{Birkedal:Category}
  and it is easy to see that locally contractive functors in $\Sl$ are locally contractive in
  the $M$-category sense. Because fixed points exist in $M$-categories, the fixed point
  of $F$ exists in $ti\Sl$.
\qed

\begin{cor}\label{cor:glambda_types_TI}
  All $\glambda$-types that do not have the empty type $\EMPTY$ in their syntax tree
  are total and inhabited.
\end{cor}
\proof
The $\mu$-case is covered by Lemma~\ref{prop:functor-restr-fp-total}, because
open types whose free variables are guarded denote locally contractive functors; the
$\blacksquare$ case holds because total and inhabited objects $X$ admit at least one
global element $1\to X$; all other cases are routine.
\qed

Further sound reasoning principles in $\logiclambdanext$, some making use of the
concept of total and inhabited type, are listed in Figure~\ref{fig:old-rules}, and in the
lemmas below, whose proofs are all routine. Note that the rule $\eqlaternextrule$
establishes a close link between $\LATER$ and $\later$, as
Lemma~\ref{prop:always-unbox-injective} does for $\blacksquare$ and $\always$.

\begin{figure}
  \centering
  \begin{mathpar}
    \inferrule*[Right=L\"ob]{ }{\Gamma \mid \Xi, (\later \phi \implies \phi) \vdash \phi}
    \and
    \inferrule*[Right=$\exists\later$]{ }{\Gamma, x : X \mid \exists y : Y, \later
      \phi(x,y) \vdash \later \left(\exists y : Y, \phi(x, y)\right)}
    \and
    \inferrule*[Right=$\forall\later$]{ }{\Gamma, x:X \mid \later (\forall y : Y, \phi(x, y)) \vdash \forall y :
      Y, \later \phi(x, y)}
    \and
    \inferrule*{ }{\Gamma \mid \Xi, \phi \vdash \later \phi}
    \and
    \inferrule*{\star \in \{\land, \lor, \implies\} }
    {\Gamma \mid \later (\phi \star \psi) \dashv\vdash \later\phi \star \later\psi}
    \and
    \inferrule*{\Gamma \mid \lnot\lnot \phi \vdash \psi}{\Gamma \mid \phi \vdash \always \psi}
    \and
    \inferrule*{\Gamma \mid \phi \vdash \always \psi}{\Gamma \mid \lnot\lnot \phi \vdash \psi}
    \and
    \inferrule*{\Gamma \mid \phi \vdash \psi}{\Gamma \mid \always \phi \vdash \always \psi}
    \and
    \inferrule*{ }{\Gamma \mid \always \phi \vdash \phi}
    \and \inferrule*{ }{\Gamma \mid \always \phi \vdash \always\always \phi}
    \and
    \inferrule*[Right=\eqlaternextrule]{ }{\forall x, y : X . \later (x =_X y) \iff \NEXT x =_{\LATER X} \NEXT y}
    \phantom{\eqlaternextrule}
  \end{mathpar}
  \caption{Valid rules for $\later$ and $\always$. The
    converse entailment in $\forall\later$ and $\exists\later$ rules holds if $Y$ is
    total and inhabited. In all rules involving $\always$ the context $\Gamma$
    is assumed constant.}
  \label{fig:old-rules}
\end{figure}

\begin{lem}
  \label{prop:theta-is-a-fp}
  For any type $A$ and term $f : \LATER A \to A$ we have
  $\Theta f =_{A} f \left(\NEXT (\Theta f)\right)$ and, if $u$ is any other term such that
  $f (\NEXT u) =_{A} u$, then $u =_{A}\Theta f$.
\qed
\end{lem}

Finally, in the next section we will come to the problem of proving
$x=_{\blacksquare A}y$ from $\UNBOX x=_A\UNBOX y$. This does not hold in general,
but using the semantics of $\logiclambdanext$ we can prove
the proposition below.

\begin{lem}
  \label{prop:always-unbox-injective}
  The formula $\always(\UNBOX x =_A \UNBOX y) \implies x =_{\blacksquare A} y$ is valid.
\qed
\end{lem}

\subsection{Examples}\label{sec:logic-exs}

In this section we see examples of $\logiclambdanext$ proofs regarding
$\lambdanext$-programs.

\begin{exa}\label{ex:logic_exs}\hfill
\begin{enumerate}
\item\label{ex:logic_exs_map}
For any $f : A \to B$ and $g : B \to C$ we have
    \begin{equation}\label{eq:map_prop}
      (\map f) \comp (\map g) =_{\gStream{A} \to \gStream{C}} \map (f \comp g).
    \end{equation}
    Equality of functions is extensional, so it suffices to show that these are equal on
    any stream of type $\gStream{A}$, for which we use the variable $s$.
    The proof proceeds by unfolding the definitions on each side, observing that the
    heads are equal, then proving equality of the tails by \emph{L\"{o}b induction}; i.e.
    our induction hypothesis will be \eqref{eq:map_prop} with $\later$ in front:
    \begin{equation}\label{eq:map_prop_lob}
      \later((\map f) \comp (\map g) = \map (f \comp g)).
    \end{equation}
    Now unfolding the left hand side of \eqref{eq:map_prop} applied to $s$, using the
    definition of $\map$ from Example~\ref{ex:programs}.\ref{ex:map}, along with
    $\beta$-rules and Lemma~\ref{prop:theta-is-a-fp}, we get
    \[
      f(g(\head s)) \consin \left(\NEXT (\map f) \APP ((\NEXT (\map g)) \APP \tail s)\right)
    \]
    By applying the composition rule for $\APP$ this simplifies to
    \[
      f(g(\head s)) \consin \left(
        (\NEXT\mathsf{comp})\APP(\NEXT(\map f)) \APP(\NEXT (\map g))\APP \tail s
      \right)
    \]
    Applying the reduction rule for $\APP$ we simplify this further to
    \begin{equation}\label{eq:map_prop1}
      f(g(\head s)) \consin \left(\NEXT ((\map f) \comp (\map g)) \APP \tail s\right)
    \end{equation}   
    Unfolding the right of \eqref{eq:map_prop} similarly, we get
    \begin{equation}\label{eq:map_prop2}
      f(g(\head s)) \consin \left(\NEXT (\map (f\comp g)) \APP \tail s\right)
    \end{equation}
    These streams have the same head; we proceed on the tail using our induction
    hypothesis \eqref{eq:map_prop_lob}. By $\eqlaternextrule$ we immediately have
    \[
      \NEXT((\map f) \comp (\map g)) = \NEXT \map (f \comp g)
    \]
    replacing equals by equals then makes \eqref{eq:map_prop1} equal to
    \eqref{eq:map_prop2}; \textsc{L\"ob} completes the proof.
\item
  We now show how $\logiclambdanext$ can prove a second-order property. Given a
  predicate $P$ on a type $A$, that is, $P:A\to\Omega$, we can lift this to a predicate
  $\liftstr{P}$ on $\gStream{A}$ expressing that $P$ holds for all elements of the
  stream by the definition
  \begin{align*}
    \liftstr{P} \defeq \Theta\lambda r . \lambda s . P (\head s) \land
    \lift \left(r \APP (\tail s)\right) \;:\; \gStream{\NN} \to \Omega
  \end{align*}
  We can now prove for a \emph{total and inhabited} type $A$ that
  \begin{align*}
    &\forall P, Q : (A \to \gCoNat), \forall f : A \to A,
    (\forall x : A, P(x) \implies Q(f(x)))\\
    &\implies\forall s : \Stream{A}, \liftstr{P}(s) \implies \liftstr{Q}(\map f\,s).
  \end{align*}
Recall that $\map$ satisfies $\map f\,s = f(\head s) \consin (\NEXT (\map f) \APP
(\tail s))$. We will prove the property by L\"ob induction, and so assume
\begin{align}
  \label{eq:lob-IH-map}
  \later(\forall s : \Stream{\NN}, \liftstr{P}(s) \implies \liftstr{Q}(\map f\,s))
\end{align}
Let $s$ be a stream satisfying $\liftstr{P}$. If we unfold $\liftstr{P}(s)$ we get
$P(\head s)$ and $\lift (\NEXT \liftstr{P} \APP (\tail s))$. We need to prove
$Q(\head (\map f\,s))$ and $\lift (\NEXT \liftstr{Q} \APP (\tail (\map f\,s)))$.
The first is easy since $Q(\head (\map f\,s)) = Q(f(\head s))$. For the second we
have $\tail (\map f\,s) = \NEXT (\map f) \APP (\tail s)$. As $A$ is total and
inhabited, $\gStream{A}$ is also by Corollary~\ref{cor:glambda_types_TI}.
Hence there is a stream $s'$ such that $\NEXT s' = \tail s$. This gives
$\tail (\map f\,s) = \NEXT (\map f s')$ and so our desired result reduces to
$\lift (\NEXT (\liftstr{Q}(\map f\, s')))$ and $\lift (\NEXT \liftstr{P} \APP (\tail
s))$ is equivalent to $\lift (\NEXT (\liftstr{P}(s')))$. But $\lift \comp \NEXT =
\later$ and so the  induction hypothesis \eqref{eq:lob-IH-map} and \textsc{L\"ob} finish
the proof.
\end{enumerate}
\end{exa}\medskip

\noindent We now turn to examples that involve the constant type-former $\blacksquare$.

\begin{exa}\label{ex:logic_const}\hfill
  \begin{enumerate}
\item
  Recall the functions $\iterate':(A\to A) \to A \to \gStream{A}$ of
  Example~\ref{ex:programs}.\ref{exa:iterate_and_interleave} and $\everysecond:
  \Stream{A}\to\gStream{A}$ of
  Example~\ref{exa:constant_progs}.\ref{exa:every2nd}. Then for every $x:A$ and
  $f:A\to A$,
    \[\everysecond (\BOX \iota . \iterate'\,f\, x)
      =_{\gStream{A}} \iterate'\, f^2\, x\]
  where $f^2$ is $\lambda x.f(fx)$.

  First we prove the intermediate result
  \begin{equation}
  \label{eq:iterate-tail}
  \limtail (\BOX \iota . \iterate'\,f\, x) =_{\Stream{A}} 
  \BOX \iota . \iterate' \, f\, (f \,x)
  \end{equation}
  which follows by:
  \begin{align*}
    \limtail\, (\BOX \iota . \iterate'\,f\, x) &=
    \BOX\, [s \explsubst \BOX \iota . \iterate'\,f\, x] . \PREV \iota . \tail\UNBOX s \\
    &= \BOX\iota.\PREV[s \explsubst \BOX \iota . \iterate'\,f\, x] . \tail\UNBOX s \\
    &= \BOX\iota.\PREV\iota.\tail\UNBOX\BOX \iota . \iterate'\,f\, x \\
    &= \BOX\iota.\PREV\iota.\tail\iterate'\,f\, x \\
    &= \BOX\iota.\PREV\iota.(\NEXT\iterate' f)\APP(\NEXT (f\,x)) \\
    &= \BOX\iota.\PREV\iota.\NEXT(\iterate' f (f\,x)) \\
    &= \BOX\iota.\iterate' f (f\,x)
  \end{align*}
  The first step follows by the definition of $\limtail$ and the $\beta$-rule for functions.
  The next two steps require the ability to move substitutions through a $\BOX$ and
  $\PREV$; see the last three equations of Figure~\ref{fig:additional-equations}. The
  remaining steps follow from unfolding definitions, various $\beta$-rules, and
  Lemma~\ref{prop:theta-is-a-fp}.

  Now for L\"ob induction assume
  \begin{equation}
    \label{eq:LIH}
    \later \left( \everysecond (\BOX \iota . \iterate'\,f\, x)
      =_{\gStream{A}} \iterate'\, f^2\, x\right),
  \end{equation}
  then we can derive
  \begin{align*}
  \everysecond\, (\BOX \iota.&\iterate'\,f\,x) \\
  &= x \consin (\NEXT\everysecond)\APP(\NEXT\limtail\limtail\BOX\iota.\iterate'\,f\,x) \\
  &= x \consin \NEXT\everysecond\limtail\limtail\BOX\iota.\iterate'\,f\,x \\
  &= x \consin \NEXT\everysecond\BOX\iota.\iterate'\,f(f^2\,x)
    \tag*{\eqref{eq:iterate-tail}} \\
  &= x \consin \NEXT\iterate' f^2\, (f^2\,x)
    \tag*{\eqref{eq:LIH} and \eqlaternextrule} \\
  &= \iterate'\, f^2\, x
  \end{align*}
  One might wonder why we use $\iterate'$ here instead of the more general $\iterate$;
  the answer is that we cannot form the subterm $\BOX \iota . \iterate\,f\, x$ if $f$ is
  a variable of type $\LATER(A\to A)$, because this is not a constant type. 
  \item\label{ex:limit_logic}
    Given a term in constant context $f:A \to B$ we define
    \[
      \mathcal{L}(f) \defeq \lim \BOX\iota.f:\blacksquare A \to \blacksquare B
    \]
    recalling $\lim$ from Example~\ref{exa:constant_progs}.\ref{exa:lim}. For any
    such $f$ and $x:\blacksquare A$ we can then prove $\UNBOX (\mathcal{L}(f)\,x)
    =_{B} f(\UNBOX x)$. This allows us to prove, for example,
    \begin{equation}\label{eq:lifting_prop}
      \mathcal{L}(f \comp g) = \mathcal{L}(f) \comp\mathcal{L}(g)
    \end{equation}
    as follows: $\UNBOX(\mathcal{L}(f \comp g)(x)) =f\comp g(\UNBOX x)=
    \UNBOX(\mathcal{L}(f) \comp\mathcal{L}(g)(x))$. This is true without any
    assumptions, and so $\always(\UNBOX(\mathcal{L}(f \comp g)(x))=
    \UNBOX(\mathcal{L}(f) \comp\mathcal{L}(g)(x)))$, so by
    Lemma~\ref{prop:always-unbox-injective} and functional extensionality,
    \eqref{eq:lifting_prop} follows.

   For functions of arity $k$ we define $\mathcal{L}_k$ using $\mathcal{L}$, and
   analogous properties hold, e.g. we have $\UNBOX(\mathcal{L}_2(f)\,x\,y) =
   f(\UNBOX x)(\UNBOX y)$, which allows us to lift equalities proved for functions on
   guarded types to functions on constant types; see
   Section~\ref{sec:definable-functions} for an example.
\item
  In Section~\ref{subsec:boxplus} we claimed there is an isomorphism between the
  types $\blacksquare A+\blacksquare B$ and $\blacksquare(A+B)$, witnessed by the
  terms
  \[
  \begin{array}{lcl}
  \lambda x.\BOX \iota.\CASE x\OF x_1.\IN_1\UNBOX x_1;x_2.\IN_2\UNBOX x_2 &:&
    (\blacksquare A+\blacksquare B)\to \blacksquare(A+B) \\
  \lambda x.\BOXSUM \iota.\UNBOX x &:&
    \blacksquare(A+B)\to\blacksquare A+\blacksquare B
  \end{array}
  \]
  We are now in a position to prove that these terms are mutually inverse. In the below
  we use the rules regarding the permutation of substitutions through $\BOXSUM$, the
  interaction of $\BOXSUM$ with $\CASE$, and $\eta$-rules for sums and
  $\blacksquare$:
  \[\begin{array}{l}
    (\lambda x.\BOX \iota.\CASE x\OF x_1.\IN_1\UNBOX x_1;x_2.\IN_2\UNBOX x_2)(\BOXSUM \iota.\UNBOX x) \\
    =\; \BOX[x\explsubst\BOXSUM \iota.\UNBOX x].\CASE x\OF x_1.\IN_1\UNBOX x_1;x_2.\IN_2\UNBOX x_2 \\
    =\; \BOX\iota.\CASE (\BOXSUM \iota.\UNBOX x)\OF x_1.\IN_1\UNBOX x_1;x_2.\IN_2\UNBOX x_2 \\
    =\; \BOX\iota.\CASE(\UNBOX x)\OF x_1.\IN_1 x_1;x_2.\IN_2 x_2 \\
    =\; \BOX\iota.\UNBOX x \\
    =\; x
  \end{array}\]
  The other direction requires the permutation of a substitution through $\BOXSUM$,
  the $\beta$-rule for $\blacksquare$, the commuting conversion of $\BOXSUM$ through
  $\CASE$, the reduction rule for $\BOXSUM$, and $\eta$-rules for $\blacksquare$ and
  sums:
  \[\begin{array}{l}
    (\lambda x.\BOXSUM \iota.\UNBOX x)(\BOX \iota.\CASE x\OF x_1.\IN_1\UNBOX x_1;x_2.\IN_2\UNBOX x_2) \\
    =\; \BOXSUM[x\explsubst\BOX \iota.\CASE x\OF x_1.\IN_1\UNBOX x_1;x_2.\IN_2\UNBOX x_2].\UNBOX x \\
    =\; \BOXSUM\iota.\UNBOX\BOX \iota.\CASE x\OF x_1.\IN_1\UNBOX x_1;x_2.\IN_2\UNBOX x_2 \\
    =\; \BOXSUM\iota.\CASE x\OF x_1.\IN_1\UNBOX x_1;x_2.\IN_2\UNBOX x_2 \\
    =\; \CASE x\OF x_1.\BOXSUM\iota.\IN_1\UNBOX x_1;x_2.\BOXSUM\iota.\IN_2\UNBOX x_2 \\
    =\; \CASE x\OF x_1.\IN_1\BOX\iota.\UNBOX x_1;x_2.\IN_2\BOX\iota.\UNBOX x_2 \\
    =\; \CASE x\OF x_1.\IN_1 x_1;x_2.\IN_2 x_2 \\
    =\; x
  \end{array}\]
  \end{enumerate}
\end{exa}\medskip

\noindent As a final remark of this section, we note that our main direction of further work beyond
this paper has been to extend the $\glambda$-calculus with dependent
types~\cite{Bizjak:Guarded}, as we will discuss further in Section~\ref{sec:further}. In
this setting proofs take place inside the calculus, as with proof assistants such as
Coq~\cite{Coq:manual} and Agda~\cite{Norell:Towards}. The `pen-and-paper' proofs
of this section are therefore interesting partly because they reveal some of the
constructions that are essential to proving properties of guarded recursive programs;
these are the constructions that must be supported by the dependent type theory.


\section{Behavioural Differential Equations}
\label{sec:definable-functions}

In this section we demonstrate the expressivity of the approach of this paper by
showing how to construct coinductive streams as solutions to \emph{behavioural
differential equations}~\cite{Rutten:2003:bde} in the $\lambdanext$-calculus. This
hence allows us to reason about such functions in $\logiclambdanext$, instead of via
bisimulation arguments.\enlargethispage{\baselineskip}

\subsection{Definition and Examples}

We now define, and give examples of, behavioural differential equations. These
examples will allow us to sketch informally how they can be expressed within the
$\glambda$-calculus, and how the program logic $\logiclambdanext$ can be used to
reason about them.

\begin{defi}\label{def:bde}
Let $\Sigma$ be a first-order signature over a base sort $A$.
A \emph{behavioural differential equation} for a $k$-ary
stream function is a pair of terms $h_f$ and $t_f$ (standing for \emph{head} and
\emph{tail}), where $h_f$ is a term containing function symbols from $\Sigma$, and
variables as follows:
\begin{align*}
  \hastype{x_1,\ldots,x_k : A}{h_f}{A}
\end{align*}
Intuitively, the variables $x_i$ denote the heads of the argument stream.
$t_f$ is a term with function symbols from $\Sigma$ along with a new constant
$f$ of sort $\left(\Stream{A}\right)^k \to \Stream{A}$, and variables as follows:
\begin{align*}
  \hastype{x_1,\ldots,x_k,y_1,\ldots,y_k,z_1,\ldots,z_k : \Stream{A}}{t_f}{\Stream{A}}
\end{align*}
Intuitively, the variables $x_i$ denote the streams whose head is the head of the
argument stream and whose tails are all zeros, the variables $y_i$ denote the argument
streams, the variables $z_i$ denote the tails of the argument streams, and the new
constant $f$ is recursive self-reference.

Further, given a set of stream functions defined by behavioural differential equations,
the term $t_f$ can use functions from that set as constants (behavioural differential
equations are therefore \emph{modular} in the sense of Milius et
al.~\cite{milius:abstract-gsos}).
\end{defi}

Note that we have slightly weakened the original notion of behavioural differential
equation by omitting the possibility of mutually recursive definitions, as used for
example to define the stream of Fibonacci numbers~\cite[Section 5]{Rutten:2003:bde}.
This omission will ease the notational burden involved in the formal results of the next
section, but mutually recursive definitions can be accommodated within the
$\glambda$-calculus setting by, for example, considering a pair of mutually recursive
stream functions as a function producing a pair of streams.

\begin{exa}\label{exa:bdes}\hfill
\begin{enumerate}
\item
  Assuming we have constant $\ZERO$ of type $\NAT$, the constant stream $\zeros$ of
  Example~\ref{ex:programs}.\ref{ex:zeros} is defined as a behavioural differential
  equation by
  \[
    h_{\zeros} \;=\; \ZERO
    \qquad\qquad
    t_{\zeros} \;=\; \zeros
  \]
\item
  As an example of the modularity of this setting, given some $n:\NAT$ we can
  define the stream $[n]$ using the $\zeros$ stream defined above, by
  \[
    h_{[n]} \;=\; n
    \qquad\qquad
    t_{[n]} \;=\; \zeros
  \]
\item\label{ex:streamplusbde}
  Assuming we have addition $+:\NAT\times\NAT\to\NAT$ written infix, then stream
  addition, also written $+$ and infix, is the binary function defined by
  \[
    h_{+} \;=\; x_1+x_2
    \qquad\qquad
    t_{+} \;=\; z_1+z_2
  \]
\item
  Assuming we have multiplication $\times:\NAT\times\NAT\to\NAT$, written infix, then
  stream product, also written $\times$ and infix, is the binary function defined by
    \[
    h_{\times} \;=\; x_1\times x_2
    \qquad\qquad
    t_{\times} \;=\; (z_1\times y_2)+(x_1\times z_2)
  \]
\end{enumerate}
\end{exa}\medskip

\noindent It is straightforward to translate the definitions above into constructions on guarded streams in the $\glambda$-calculus. For example, stream addition is defined by the
function on guarded streams
$\plus : \gStream{\NAT} \to \gStream{\NAT} \to \gStream{\NAT}$ below:
\begin{align*}
   \plus \defeq 
   \Theta \lambda p . \lambda s_1 . \lambda s_2 . (\head s_1 + \head s_2)\consin(p \APP (\tail  s_1) \APP (\tail s_2))
\end{align*}
We can lift this to a function on streams $\limplus : \Stream{\NAT}\to \Stream{\NAT}
\to\Stream{\NAT}$ by $\limplus\defeq\mathcal{L}_2(\plus)$, recalling $\mathcal{L}_2$
from Example~\ref{ex:logic_const}.\ref{ex:limit_logic}. Now by
Lemma~\ref{prop:theta-is-a-fp} we have
\begin{align}
  \label{eq:plus-defining-eq}
 \plus = \lambda s_1 . \lambda s_2 . (\head s_1 + \head s_2)\consin ((\NEXT\plus) \APP (\tail s_1)
  \APP (\tail s_2)).
\end{align}
We can then prove in the logic $\logiclambdanext$ that the definition of $\limplus$
satisfies the specification given by the behavioural differential equation of
Example~\ref{exa:bdes}.\ref{ex:streamplusbde}.
Given $s_1,s_2: \Stream{\NAT}$, we have
\[\begin{array}{rcll}
  \limhead(\limplus s_1 s_2) &=& \head\UNBOX(\limplus s_1\,s_2) \\
  &=& \head\UNBOX(\mathcal{L}_2(\plus)\,s_1\,s_2) \\
  &=& \head(\plus(\UNBOX s_1)(\UNBOX s_2)) & \mbox{(Example~\ref{ex:logic_const}.\ref{ex:limit_logic})} \\
  &=& (\head\UNBOX s_1)+(\head\UNBOX s_2) & \eqref{eq:plus-defining-eq} \\
  &=& (\limhead s_1)+(\limhead s_2)
\end{array}\]
For the $\limtail$ case, that $\limtail(\limplus s_1\,s_2)=\limplus(\limtail s_1)
(\limtail s_2)$, we proceed similarly, but also using that $\tail (\UNBOX \sigma) =
\NEXT (\UNBOX(\limtail \sigma))$ which follows from the definition of $\limtail$, the
$\beta$-rule for $\blacksquare$, and the $\eta$-rule for $\LATER$.

We can hence use $\logiclambdanext$ to prove further properties of streams defined
via behavioural differential equations, for example that stream addition is commutative.
Such proofs proceed by conducting the proof on the guarded stream produced by
applying $\UNBOX$, then by introducing the $\always$ modality so long as the context
is suitably constant, and then by invoking Lemma~\ref{prop:theta-is-a-fp}.

\subsection{From Behavioural Differential Equations to $\glambda$-Terms}
\label{sec:from-bde-to-terms}

In the previous section we saw an example of a translation from a behavioural
differential equation to a $\glambda$-term. In this section we present the general
translation. Starting with a $k$-ary behavioural differential equation
$(h_f,t_f)$ we will define a $\glambda$-term\footnote{We use the uncurried form to simplify the semantics.}
\begin{align*}
  \guarded{\Phi_f} :
  \LATER \left(\left(\gStream{A}\right)^k \to \gStream{A}\right)
  \to \left(\left(\gStream{A}\right)^k \to \gStream{A}\right)
\end{align*}
by induction on the structure of $h_f$ and $t_f$. We may apply a fixed-point
combinator to this to get a function on guarded streams, which we write as
$\guarded{f}$.

We first extend $\glambda$ with function symbols in the signature $\Sigma$ of $(h_f,t_f)$.
Using these it is straightforward to define a $\glambda$ term $\guarded{h_f}$ of type
\begin{align*}
  x_1 : A, x_2 : A, \cdots, x_k : A \vdash \guarded{h_f} : A,
\end{align*}
corresponding to $h_f$ in the obvious way.

From $t_f$ we define the term $\guarded{t_f}$ of type
\begin{align*}
  \hastype{\vec{x}, \vec{y} : \gStream{A}, \vec{z} : \LATER\gStream{A},
  f : \LATER \left(\left(\gStream{A}\right)^k \to \gStream{A}\right)}{\guarded{t_f}}{\LATER\gStream{A}}
\end{align*}
by induction on the structure of $t_f$ as follows.

The base cases are simple:\enlargethispage{\baselineskip}
\begin{itemize}
\item If $t_f = x_i$ for some $i$ we put $\guarded{t_f} = \NEXT x_i$, and similarly for
  $y_i$;
\item If $t_f = z_i$ we put $\guarded{t_f} = z_i$.
\end{itemize}

If $t_f = f(a_1,\ldots, a_k)$ we put
\begin{align*}
  \guarded{t_f} = \guarded{\operatorname{curry}}(f) \APP \guarded{t_{a_1}} \APP \cdots \APP \guarded{t_{a_k}}
\end{align*}
where $\guarded{\operatorname{curry}}(f)$ is the currying of the function $f$, which is easily definable as a $\glambda$ term.

Finally if $t_f = e(a_1,\ldots, a_l)$ for some previously defined $l$-ary $e$ then we put
\begin{align*}
  \guarded{t_f} = \guarded{\operatorname{curry}}(\NEXT \guarded{e}) \APP \guarded{t_{a_1}} \APP \cdots \APP
  \guarded{t_{a_l}}
\end{align*}
We can then combine the terms $\guarded{h_f}$ and $\guarded{t_f}$ to define the
desired term $\guarded{\Phi_f}$ as
\[
  \lambda f,\vec{y}.(\guarded{h_f}[\head y_i/x_i])\consin
  (\guarded{t_f}[(\head y_i\consin\NEXT\zeros)/x_i,\tail y_i/z_i])
\]
Analogously from a behavioural differential equation we define a $\glambda$ term $\Phi_f$ of type
\begin{align*}
  \Phi_f : \left(\left(\Stream{A}\right)^k \to \Stream{A}\right) \to \left(\left(\Stream{A}\right)^k \to \Stream{A}\right),
\end{align*}
where for the function symbols we take the lifted (as in
Example~\ref{ex:logic_const}.\ref{ex:limit_logic}) function symbols used in the definition
of $\guarded{\Phi_f}$.

We will now show that the lifting of the unique fixed point of $\guarded{\Phi_f}$ is a fixed point of $\Phi_f$, and hence satisfies the behavioural differential equation for $f$.
We prove this using denotational semantics, relying on its adequacy (Corollary~\ref{cor:adequacy:den-implies-ctxeq}).

\subsection{The Topos of Trees as a Sheaf Category}\label{sec:sheaf}

In order to reach the formal results regarding behavioural differential equations of the
next section, it will be convenient to provide an alternative definition for the topos of
trees as a category of \emph{sheaves}, rather than \emph{presheaves}.

The preorder $\omega=1\leq 2\leq\cdots$ is a topological space given the
\emph{Alexandrov topology} where the open sets are the \emph{downwards} closed sets. These
downwards closed sets are simply $0\subseteq 1\subseteq 2\subseteq\cdots\subseteq
\omega$, where $0$ is the empty set, $n$ is the downwards closure of $n$ for any
positive integer $n$, and $\omega$ is the entire set. Then the sheaves $X$
over this topological space, $\Sh{\omega}$, are presheaves over these open sets
obeying certain properties~\cite{Maclane:Sheaves}. In this case these properties
ensure that $X(0)$ must always be a singleton set and $X(\omega)$ is entirely
determined (up to isomorphism) by the sets $X_1,X_2,\cdots$ as their \emph{limit}.
This definition is hence plainly equivalent to the definition of $\trees$ from
Section~\ref{sec:tot}.

However this presentation is more convenient for our purposes here, in which we will
need to go back and forth between the categories $\trees$ and $\sets$, because the
global sections functor%
\footnote{The standard notation $\Gamma$ for this functor should not be confused with
our notation for typing contexts.}
$\Gamma$ in the sequence of adjoints
\begin{align*}
  \Pi_1 \dashv \Delta \dashv \Gamma
\end{align*}
where
\begin{align*}
  \begin{split}
    \Pi_1 &: \Sl \to \sets\\
    \Pi_1(X) &= X(1)
  \end{split}
  \qquad
  \begin{split}
    \Delta &: \sets \to \Sl\\
    \Delta(a)(\alpha) &=
    \begin{cases}
      1 & \text{if } \alpha = 0\\
      a & \text{otherwise}
    \end{cases}
  \end{split}
  \qquad
  \begin{split}
    \Gamma &: \Sl \to \sets\\
    \Gamma(X) &= X(\omega)
  \end{split}
\end{align*}
is just evaluation at $\omega$, i.e. the limit is already present, which simplifies
notation. Another advantage is that $\LATER : \Sl \to \Sl$ is given as
\begin{align*}
  (\LATER X)(\nu+1) &= X(\nu)\\
  (\LATER X)(\alpha) &= X(\alpha)
\end{align*}
where $\alpha$ is a limit ordinal (either $0$ or $\omega$) which means that $\LATER
X(\omega) = X(\omega)$ and as a consequence, $\nxt_{\omega} = \id{X(\omega)}$
and $\Gamma(\LATER X) = \Gamma(X)$ for any $X \in \Sl$ and so $\blacksquare
(\LATER X) =\blacksquare X$ for any $X$, so we do not have to deal with mediating
isomorphisms.

We finally turn to a useful lemma which we will use in the next section.

\begin{lem}
  \label{lem:fixed-point-mapped-to-fp}
  Let $X, Y$ be objects of $\Sl$. Let $F : \LATER\left(Y^X\right) \to Y^X$ be a
  morphism in $\Sl$ and $\underline{F}:Y(\omega)^{X(\omega)}\to
  Y(\omega)^{X(\omega)}$ be a \emph{function} in $\sets$. Suppose that the diagram
  \[
  \begin{largediagram}
    \Gamma\left(\LATER\left(Y^X\right)\right) \ar{r}[description]{\Gamma(F)} \ar{d}[description]{\lim} & \Gamma(Y^X) \ar{d}[description]{\lim}\\
    Y(\omega)^{X(\omega)} \ar{r}[description]{\underline F} & Y(\omega)^{X(\omega)}
  \end{largediagram}
  \]
  commutes, where $\lim\left(\{g_\nu\}_{\nu=0}^{\omega}\right) = g_{\omega}$. By
  Banach's fixed point theorem $F$ has a unique fixed point, say $u : 1 \to Y^X$.

  Then $\lim(\Gamma(u)(\ast)) = \lim(\Gamma(\nxt \comp u)(\ast)) =
  \Gamma(\nxt \comp u)(\ast)_{\omega} = u_{\omega}(\ast)_{\omega}$ is a fixed point of $\underline F$.
\end{lem}
\proof
  \begin{align*}
    \underline F \left(\lim(\Gamma(u)(\ast))\right) &\;=\;
    \lim(\Gamma(F)(\Gamma(\nxt \comp u)(\ast)))\\
    &\;=\; \lim(\Gamma(F \comp \nxt \comp u)(\ast)) \;=\;
    \lim(\Gamma(u)(\ast)).\rlap{\hbox to 71 pt{\hfill\qEd}}
  \end{align*}\smallskip

\noindent Note that $\lim$ is not an isomorphism, as there are in general many more functions
from $X(\omega)$ to $Y(\omega)$ than those that arise from natural transformations.
The ones that arise from natural transformations are the \emph{non-expansive} ones.

\subsection{Expressing Behavioural Differential Equations}
\label{sec:rutt-behav-diff}

We first define two interpretations of behavioural differential equations (Definition~\ref{def:bde}); first in the topos of trees, and then in $\sets$.
The interpretation in $\Sl$ is just the denotation of the term $\guarded{\Phi_f}$ from Section~\ref{sec:from-bde-to-terms}, whereas the inclusion of the interpretation in $\sets$ into the topos of trees, using the constant presheaf functor $\Delta$, is the denotation of the term $\Phi_f$ from Section~\ref{sec:from-bde-to-terms}.

\begin{defi}
Fixing a set $|A|$ which will interpret our base sort, define $\denS{A} = \Delta|A|$
and $\denS{\Stream{A}} =\mu X . \Delta|A| \times \LATER X$; that is, the denotation
of $\gStream{(\Delta|A|)}$ from Example~\ref{ex:denote_streams}.\ref{ex:gStream}.
To each function symbol $g \in \Sigma$ of type
$\tau_1, \ldots, \tau_n \to \tau_{n+1}$ we assign a morphism
\begin{align*}
  \denS{g} : \denS{\tau_1} \times \denS{\tau_2} \times \cdots \times \denS{\tau_n} \to
  \denS{\tau_{n+1}}.
\end{align*}
We then interpret $h_f$ as a morphism of type $\denS{A}^k \to \denS{A}$ by
induction:
\begin{align*}
  \denS{x_i} &= \pi_i\\
  \denS{g(t_1, t_2, \ldots, t_n)} &= \denS{g} \comp \left\langle \denS{t_1}, \denS{t_2},
    \cdots, \denS{t_n}\right\rangle.
\end{align*}
$t_f$ will be interpreted similarly, but we also have the new function symbol $f$ to
consider. The interpretation of $t_f$ is therefore
a $\trees$-arrow of type
\begin{align*}
  \denS{t_f} : \denS{\Stream{A}}^k \times \denS{\Stream{A}}^k \times 
  \left(\LATER\left(\denS{\Stream{A}}\right)\right)^k \times
  \LATER\left(\denS{\Stream{A}}^{\denS{\Stream{A}}^k}\right) \to
  \LATER(\denS{\Stream{A}})
\end{align*}
and is defined as:
\begin{align*}
  \denS{x_i} &= \nxt \comp \pi_{x_i}\\
  \denS{y_i} &= \nxt \comp \pi_{y_i}\\
  \denS{z_i} &= \pi_{z_i}\\
  \denS{g(t_1, t_2, \ldots, t_n)} &= \LATER(\denS{g}) \comp \mathbf{can} \comp
  \left\langle \denS{t_1}, \denS{t_2},\cdots, \denS{t_n}\right\rangle &\text{if } g \neq f\\
  \denS{f(t_1, t_2, \ldots, t_k)} &=
  \mathbf{eval} \comp \left\langle J \comp \pi_f,
    \mathbf{can} \comp \left\langle \denS{t_1}, \denS{t_2}, \cdots, \denS{t_k}\right\rangle\right\rangle
\end{align*}
where $\mathbf{can}$ is the canonical isomorphism witnessing that $\LATER$ preserves products; $\mathbf{eval}$ is the evaluation map, and $J$ is the map $\LATER(X\to Y)\to\LATER X\to\LATER Y$ which gives $\LATER$ its applicative functor structure $\APP$.

We can then define the $\trees$-arrow
\[
F : \LATER\left(\denS{\Stream{A}}^{\denS{\Stream{A}}^k}\right) \to
\denS{\Stream{A}}^{\denS{\Stream{A}}^k}
\]
as the exponential transpose of
\begin{align*}
  F' = \mathbf{fold} \comp \left\langle \den{h_f} \comp \vec{\mathbf{hd}} \comp \pi_1, \denS{t_f} \comp
  \left(\left\langle \vec{\iota \comp \mathbf{hd}}, \id{\denS{\Stream{A}}^k}, \vec{\mathbf{tail}}\right\rangle\times
\id{\LATER\left(\denS{\Stream{A}}^{\denS{\Stream{A}}^k}\right)} \right)\right\rangle
\end{align*}
where $\mathbf{hd}$ and $\mathbf{tl}$ are head and tail functions, extended in the obvious way to tuples, and $\mathbf{fold}:\denS{A}\times\LATER\denS{\Stream{A}}\to\denS{\Stream{A}}$ is the evident `cons' arrow.
The function $\iota$ maps an element in $A$ to the guarded stream with head $a$ and tail the stream of zeroes.
\end{defi}

\begin{defi}
We now use the topos of trees definition above to define the denotation of $h_f$ and
$t_f$ in $\sets$. We set $\denSet{A}=|A|$ and $\denSet{\Stream{A}} =
\denS{\Stream{A}}(\omega)$. For each function symbol in $\Sigma$ we define
$\denSet{g} = \Gamma{\denS{g}} =\left(\denS{g}\right)_{\omega}$.

We then define $\denSet{h_f}$ as a function
\[\denSet{A}^k \to \denSet{A}\]
exactly as we defined $\denS{h_f}$:
\begin{align*}
  \denSet{x_i} &= \pi_i\\
  \denSet{g(t_1, t_2, \ldots, t_n)} &= \denSet{g} \comp \left\langle \denSet{t_1}, \denSet{t_2},
    \cdots, \denSet{t_n}\right\rangle.
\end{align*}
The denotation of $t_f$ is somewhat different, as we do not have the functor $\LATER$.
We define
\begin{align*}
  \denSet{t_f} : \denSet{A}^k \times  
  \denSet{\Stream{A}}^k \times \denSet{\Stream{A}}^k \times
  \denSet{\Stream{A}}^{\denSet{\Stream{A}}^k} \to \denSet{\Stream{A}}
\end{align*}
as follows:
\begin{align*}
  \denSet{x_i} &= \pi_{x_i}\\
  \denSet{y_i} &= \pi_{y_i}\\
  \denSet{z_i} &= \pi_{z_i}\\
  \denSet{g(t_1, t_2, \ldots, t_n)} &= \denSet{g} \comp
  \left\langle \denSet{t_1}, \denSet{t_2},\cdots, \denSet{t_n}\right\rangle &\text{if } g \neq f\\
  \denSet{f(t_1, t_2, \ldots, t_k)} &=
  \mathbf{eval} \comp \left\langle \pi_f,
    \left\langle \denSet{t_1}, \denSet{t_2}, \cdots, \denSet{t_k}\right\rangle\right\rangle.
\end{align*}\enlargethispage{2\baselineskip}%
We then define
\[
\underline F : \denSet{\Stream{A}}^{\denSet{\Stream{A}}^k} \to
\denSet{\Stream{A}}^{\denSet{\Stream{A}}^k}
\]
as
\[
  \underline F(\phi)\left(\vec{\sigma}\right) = \Gamma\left(\mathbf{fold}\right)\left(
  \denSet{h_f}\comp\vec{\mathbf{hd}}(\vec{\sigma}), \denSet{t_f}\left(
    \iota\left(\vec{\mathbf{hd}}(\vec{\sigma})\right), \vec{\sigma}, \vec{\mathbf{tl}}(\vec{\sigma}),\phi\right)\right)
\]
\end{defi}

\begin{lem}
  \label{prop:stream-diagram-commutes}
  For the above defined $F$ and $\underline F$ we have
  \begin{align*}
    \lim \comp \Gamma(F) = \underline F \comp \lim.
  \end{align*}
\end{lem}
\proof
  Take $\phi \in
  \Gamma\left(\LATER\left(\denS{\Stream{A}}^{\denS{\Stream{A}}^k}\right)\right) =
  \Gamma\left(\denS{\Stream{A}}^{\denS{\Stream{A}}^k}\right)$.
  We have
  \begin{align*}
    \lim(\Gamma(F)(\phi)) = \lim\left(F_{\omega}(\phi)\right) = F_{\omega}(\phi)_\omega
  \end{align*}
  and
  \begin{align*}
    \underline F (\lim(\phi)) = \underline F \left(\phi_{\omega}\right)
  \end{align*}
  These are both elements of $\denSet{\Stream{A}}^{\denSet{\Stream{A}}^k}$,
  and so are functions in $\sets$, so to show they are equal we can use elements. Take
  $\vec{\sigma} \in \denSet{\Stream{A}}^k$.
  We are then required to show
  \begin{align*}
    \underline F \left(\phi_{\omega}\right)(\vec{\sigma}) = F_{\omega}(\phi)_\omega(\vec{\sigma})
  \end{align*}
  Recall that $F$ is the exponential transpose of $F'$, so 
  $F_{\omega}(\phi)_\omega(\vec{\sigma}) = F'_{\omega}(\phi, \vec{\sigma})$. Now
  recall that composition in $\Sl$ is just composition of functions at each stage, that
  products in $\Sl$ are defined pointwise, and that $\nxt_{\omega}$ is the identity
  function.  Moreover, the $\trees$-arrow $\mathbf{hd}$ gets mapped by $\Gamma$ to
  $\mathbf{hd}$ in $\sets$ and the same holds for $\mathbf{tl}$. For the latter it is
  important that $\Gamma(\LATER(X)) = \Gamma(X)$ for any $X$.

  We thus get
  \begin{align*}
    F'_{\omega}(\phi, \vec{\sigma}) = \mathbf{fold}_{\omega}\left((\denS{h_f})_{\omega}
    \left(\mathbf{hd}(\vec{\sigma})\right), \left(\denS{t_f}\right)_{\omega}\left(\phi,
      \iota\left(\mathbf{hd}(\vec{\sigma})\right), \vec{\sigma}, \mathbf{tl}(\vec{\sigma})\right)\right)
  \end{align*}
  and also  
  \begin{align*}
    \underline F \left(\phi_{\omega}\right)(\vec{\sigma}) =
    \mathbf{fold}_{\omega}\left(\denSet{h_f}\left(\mathbf{hd}\left(\vec{\sigma}\right)\right),
      \left(\denSet{t_f}\right)\left(\phi_{\omega}, \iota\left(\mathbf{hd}(\vec{\sigma})\right), \vec{\sigma},
        \mathbf{tl}(\vec{\sigma})\right)\right)
  \end{align*}
  It is now easy to see that these two are equal, by induction on the
  structure of $h_f$ and $t_f$. The variable cases are trivial, but crucially use the fact
  that $\nxt_{\omega}$ is the identity. The cases for function symbols in $\Sigma$ are
  trivial by the definition of their denotations in $\sets$. The case
  for $f$ goes through similarly since application at $\omega$ only uses $\phi$ at
  $\omega$.
\qed

\begin{thm}
  \label{thm:diff-equations-fp}
  Let $\Sigma$ be a signature and suppose we have an interpretation in $\trees$.  Let
  $(h_f,t_f)$ be a behavioural differential equation defining a $k$-ary function $f$ using
  function symbols in $\Sigma$. The right-hand sides of $h_f$ and $t_f$ define a
  $\glambda$-term $\guarded{\Phi_f}$ of type
  \begin{align*}
    \guarded{\Phi_f} :
    \LATER \left(\gStream{\NAT}^k \to \gStream{\NAT}\right)
    \to \left(\gStream{\NAT}^k \to \gStream{\NAT}\right)
  \end{align*}
  and a term $\Phi_f$ of type
  \begin{align*}
    \Phi_f : 
    \LATER\left(\Stream{\NAT}^k \to \Stream{\NAT}\right)
    \to \left(\Stream{\NAT}^k \to \Stream{\NAT}\right)
  \end{align*}
  (here we must `lift' the interpretations of the function symbols in $\Sigma$ from
  guarded recursive streams to coinductive streams; this can be done by analogy with
  the $\mathcal{L}$ functions of Example~\ref{ex:logic_const}.\ref{ex:limit_logic}.)
  
  Let $\guarded{f} = \Theta\guarded{\Phi_f}$ be the fixed point of $\guarded{\Phi_f}$.
  Then $f = \mathcal{L}_k(\BOX \guarded f)$ is a fixed point of $\Phi_f$ which in turn
  implies that it satisfies equations $h_f$ and $t_f$.
\end{thm}
\proof
  The morphism $F$ in Lemma~\ref{lem:fixed-point-mapped-to-fp} is the interpretation of the term
  $\guarded{\Phi_f}$ from Section~\ref{sec:from-bde-to-terms}.
  The inclusion of the morphism $\underline{F}$ in Lemma~\ref{lem:fixed-point-mapped-to-fp} is the denotation of the term $\Phi_f$.
  Further, the inclusion (with $\Delta$) of the fixed point constructed in Lemma~\ref{lem:fixed-point-mapped-to-fp} is the denotation of $f$.
  
  Proposition~\ref{prop:stream-diagram-commutes} concludes the proof that $\den{f}$ is indeed a fixed point of $\den{\Phi_f}$.
  Hence by adequacy of the denotational semantics we have that $f$ is a fixed point of $\Phi_f$.
\qed

This concludes our proof that for each behavioural differential equation that defines a
function on streams, we can use the $\glambda$-calculus to define its solution.

\section{Concluding Remarks}\label{sec:conc}

We have seen how the guarded lambda-calculus, or $\glambda$-calculus, allows us to
program with guarded recursive and coinductive data structures while retaining
normalisation and productivity, and how the topos of trees provides adequate
semantics and an internal logic $\logiclambdanext$ for reasoning about
$\glambda$-programs. We have demonstrated our approach's expressivity by showing
that it can express behavioural differential equations, a well-known format for the
definition of stream functions. We conclude by surveying some related work and
discussing some future directions.

\subsection{Related Work}\label{sec:related}\hfill

\paragraph{\emph{\textbf{Other Calculi with Later.}}} Since Nakano's original
paper~\cite{Nakano:Modality} there have been a number of calculi presented
that utilise the later modality. Many of these calculi are \emph{causal}~\cite{Krishnaswami:Ultrametric,Krishnaswami:Semantic,Pottier:Typed,Krishnaswami:Higher12,Severi:Pure,Krishnaswami:Higher,Abel:Formalized},
in that they cannot express acausal but productive functions, and are therefore less
expressive in this respect than the guarded $\lambda$-calculus.
Note that this restriction is a feature, rather than a defect, for some applications
such as functional reactive programming~\cite{Krishnaswami:Ultrametric}, where
programs should indeed be prevented from reacting to an event before it has occurred.
We could similarly program in the fragment of the $\glambda$-calculus without
$\blacksquare$ to retain this guarantee. We further note that the
$\glambda$-calculus is intended to extend the
simply typed $\lambda$-calculus in as modest a way as possible while gaining the
expressivity we desire, and so we have avoided exotic features such as Nakano's
subtyping and first-class type equalities (which make type inference a non-trivial
open problem~\cite[Section 9]{Rowe:Semantic}), or the use of natural numbers to
stratify typing judgments~\cite{Krishnaswami:Ultrametric}, or
reduction~\cite{Abel:Formalized}.

Atkey and McBride's clock quantifiers~\cite{Atkey:Productive} showed how to express
\emph{acausal} functions in a calculus with later. This was extended to dependent types
by M{\o}gelberg~\cite{Mogelberg:tt-productive-coprogramming}, with improvements
made subsequently by Bizjak and M{\o}gelberg~\cite{Bizjak:Model}. However the 
conference version of this paper~\cite{Clouston:Programming} is the first to present
operational semantics for such a calculus.

Clock quantifiers differ in two main ways from this paper's use of the
modality $\blacksquare$. First, multiple clocks are useful for expressing nested
coinductive types that intuitively vary on multiple independent time streams, such as
infinite-breadth infinite-depth trees. We conjecture that we could accommodate this
by extending our calculus with multiple versions of our type- and term-formers:
$\mu^{\kappa},\LATER^{\kappa},\blacksquare^{\kappa},\NEXT^{\kappa}$ and so
forth, labelled by clocks $\kappa$. Guardedness and constantness side-conditions on
type- and term-formation would then check only the clock under consideration.
Semantics could be given via presheaves over $\omega^n$, where
$n$ is the number of clocks. One slightly awkward note is that we
appear to need a new term-former to construct the isomorphism
$\blacksquare^{\kappa}\LATER^{\kappa'}A\to\LATER^{\kappa'}\blacksquare^{\kappa}
A$, given as a first-class type equality by Atkey and
McBride~\cite{Atkey:Productive} (the other direction of this isomorphism, and the
permutation of $\blacksquare^{\kappa}$ with $\blacksquare^{\kappa'}$, are readily
definable as terms).

Second, and more importantly, clock quantifiers remove the need for term-formers
such as $\BOX$ to carry explicit substitutions. There is no free lunch however, as we
must instead handle side-conditions asserting that given clock variables are free in
the clock context; while such `freshness' conditions are common in formal calculi they
are a notorious source of error when reasoning about syntax.
Further, if explicit substitutions are to be completely avoided the
$\PREV$ constructor needs to be reworked, for example by replacing it with a
$\mathsf{force}$ term-former~\cite{Atkey:Productive}, and so we no longer have a
conventional destructor for $\LATER$, so $\beta\eta$-equalities become more complex.
Reiterating our remarks of Section~\ref{sec:untyped} we note that, with respect to
programming with the $\glambda$-calculus, the burden presented by the explicit substitutions seems
quite small, as all example programs involve identity substitutions only. Therefore our
use of the $\blacksquare$ modality seems the simpler choice, especially as it allows us
to adapt previously published work on term calculus for the modal logic Intuitionistic
S4~\cite{Bierman:Intuitionistic}. However in our work on extending guarded type theory
to dependent types~\cite{Bizjak:Guarded} the explicit substitutions become more
burdensome, resulting in our adoption of clock quantifiers for that work.

\paragraph{\emph{\textbf{Dual Contexts.}}} Our development draws extensively on
the term calculus for Intuitionistic S4 of Bierman and de
Paiva~\cite{Bierman:Intuitionistic}. Subsequent work by Davies and
Pfenning~\cite{Davies:Staged} modified Bierman and de Paiva's calculus,
removing the explicit substitutions attached to the $\BOX$ term-former. As ever there is
no free lunch, as instead a `dual context' is used -- the variable context has two
compartments, one of which is reserved for constant types. The calculus is then closed
under substitution via a modification of the definition of substitution to depend on
which context the variable is drawn from. Because, as stated above, we found the
burden of explicit substitutions not so great, we preferred to use the Bierman-de Paiva
calculus as our basis rather than deal with this more complicated notion of
substitution; however from our point of view these differences are relatively marginal
and largely a matter of taste.

\paragraph{\emph{\textbf{Ultrametric Spaces.}}} As noted in the proof of
Lemma~\ref{prop:functor-restr-fp-total}, the category $\Ml$ of bisected complete
non-empty ultrametric spaces is a complete subcategory of the topos of trees,
corresponding to the total and inhabited
$\trees$-objects~\cite[Section 5]{Birkedal-et-al:topos-of-trees}. This category $\Ml$
was shown to provide semantics for Nakano's calculus by Birkedal et
al.~\cite{Birkedal:Metric}, as well as for a related calculus with later by Krishnaswami
and Benton~\cite{Krishnaswami:Ultrametric}. These works do not feature the
$\blacksquare$ modality, but its definition is easy - it maps any space to the space with
the same underlying set, but the discrete metric. Why, then, do we instead use the
topos of trees? First, $\Ml$ is not a topos, and therefore our work reasoning with the
internal logic would not be possible. Second, $\Ml$ contains only \emph{non-empty}
spaces and so cannot model the $\EMPTY$ type. If the empty space is added then
$\LATER$ becomes undefinable: either $\LATER 0$ has underlying set $\emptyset$, in
which case there exists a map $\LATER 0\to 0$ and so the fixpoint function $(\LATER
0\to 0)\to 0$ cannot exist without creating an inhabitant of $0$, or the underlying set is 
not empty, in which case there is no map $\blacksquare \LATER 0\to\blacksquare 0$,
and so we cannot eliminate $\LATER$ in constant contexts.

\paragraph{\emph{\textbf{Sized Types.}}} The best developed type-based method for
ensuring productivity are sized types, introduced by Hughes et
al.~in 1996~\cite{Hughes:Proving}. They have now been implemented in the proof
assistant Agda, following work by Abel~\cite{Abel:MiniAgda}.
There is as yet no equivalent development employing the later modality, so direct
comparison on realistic examples with respect to criteria such as ease of use are
probably premature. However we can make some preliminary observations. First, 
defining denotational semantics in a topos was essential to the
development of the program logic $\logiclambdanext$; to our knowledge there is no
semantics of sized types yet developed that would support a similar development.
Second, the later modality has applications that appear quite unrelated to sized
types, in particular for modelling and reasoning about programming languages, 
starting with Appel et al.~\cite{Appel:Very} and including, for
example, the program logic iCAP~\cite{Svendsen:Impredicative}.
These applications require recursive types with negative occurrences of
the recursion variable, and so lie outside the scope of sized types.
The implementation of guarded recursive types directly in
a proof assistant should support such applications.
Here the most relevant comparison will be with the Coq formalisations of semantics for
later (in these cases, ultrametric semantics)~\cite{Sieczkowski:ModuRes,Jung:Higher}
as a basis for program logics. The Coq formalisation of the topos of trees via `forcing'
of Jaber et al.~\cite{Jaber:Extending} may also be useable for such reasoning. Our
hope is that implementing guarded recursive types as primitive might reduce the
overhead involved in working indirectly on encoded semantics.

\paragraph{\emph{\textbf{Similar Type- and Term-Formers}}} We finally mention two
further constructions that bear some resemblance to those of this paper. First, the
$\infty$ type-former, and `delay' $\sharp$, and `force' $\flat$ type-formers, for
coinduction in Agda~\cite[Section 2.2]{Danielsson:Subtyping}, look somewhat like
$\LATER$, $\NEXT$, and $\PREV$ respectively, but are not intended to replace syntactic
guardedness checking and so the resemblance is largely superficial. Second, the `next'
and `globally' modalities of (discrete time) Linear Temporal Logic, recently employed
as type-formers for functional reactive programming by Jeltsch~\cite{Jeltsch:Towards}
and Jeffrey~\cite{Jeffrey:LTL}, look somewhat like $\LATER$ and $\blacksquare$, but
we as yet see no obvious formal links between these approaches.

\subsection{Further Work}\label{sec:further}\hfill

\paragraph{\emph{\textbf{Dependent Types.}}}
As discussed earlier, a major goal of this research is to extend the simply-typed
$\glambda$-calculus to a calculus with dependent types. This could provide a
basis for interactive theorem proving with the later modality, integrating the sorts of
proofs we performed in Section~\ref{sec:logic} into the calculus itself. In Bizjak et
al.~\cite{Bizjak:Guarded} we have developed an extensional guarded dependent
type theory, which is proved sound in a model based on the topos of trees. This
extension is not entirely straightforward, most notably requiring novel constructions to
generalise applicative functor structure to dependent types. The next challenge is to
develop a type theory with decidable type checking, which would provide a basis for
implementation. We have developed a type theory with later~\cite{Birkedal:Cubical}
based on \emph{cubical type theory}~\cite{Cohen:Cubical}, which has a notion of path
equality which seems to interact better with the new constructs of guarded type theory
than the ordinary Martin-L\"of identity type. We conjecture that our new type theory has
decidable type checking, but this property is still open even for cubical type theory
without guarded recursion.

\paragraph{\emph{\textbf{Inference of $\glambda$ Type- and Term-Formers.}}} The
examples in this paper make clear that programming in the $\glambda$-calculus
is usually a matter of `decorating' conventional programs with our novel type- and
term-formers such as $\LATER$ and $\NEXT$. This decoration process is often
straightforward, but we are not insensitive to the burden on the programmer of
demanding large amounts of novel notation be applied to their program before it will
type-check. It would therefore be helpful to investigate algorithmic support for
automatically performing this decoration process.

\paragraph{\emph{\textbf{Full Abstraction.}}}
Corollary~\ref{cor:adequacy:den-implies-ctxeq} established the soundness of our
denotational semantics with respect to contextual equivalence. Its converse, full
abstraction, is left open. A proof of full abstraction, or a counter-example, would help us
to understand how good a model the topos of trees
provides for the $\glambda$-calculus, with respect to whether it differentiates terms
that are operationally equivalent. Conversely, if full abstraction were found to fail we
could ask whether a language extension is possible which brings the
$\glambda$-calculus closer to its intended semantics.

\section*{Acknowledgements}
We gratefully acknowledge our discussions with Andreas Abel, Robbert Krebbers, Tadeusz Litak,
Stefan Milius, Rasmus M{\o}gelberg, Filip Sieczkowski, Bas Spitters, and Andrea Vezzosi,
and the comments of the anoymous reviewers of both this paper and its conference version.
This research was supported in part by the \mbox{ModuRes} Sapere Aude Advanced Grant from The Danish Council for Independent Research for the Natural Sciences (FNU). Ale\v{s} Bizjak is supported in part by a Microsoft Research PhD grant.


\bibliography{bibl}

\end{document}

%% file: macros.tex
\newcommand{\logiclambdanext}{\ensuremath{L\mathsf{g}\lambda}}

\newcommand{\glambda}{\ensuremath{\mathsf{g}\lambda}}
\newcommand{\lambdanext}{\glambda}
\newcommand{\logicglambda}{\ensuremath{L\mathsf{g}\lambda}}
\newcommand{\NEXT}{\operatorname{\mathsf{next}}}
\newcommand{\LATER}{{\blacktriangleright}}
\newcommand{\UNFOLD}{\operatorname{\mathsf{unfold}}}
\newcommand{\FOLD}{\operatorname{\mathsf{fold}}}
\newcommand{\ABORT}{\operatorname{\mathsf{abort}}}
\newcommand{\UNIT}{\operatorname{\langle\rangle}}
\newcommand{\THEN}{\operatorname{\mathsf{then}}}
\newcommand{\ELSE}{\operatorname{\mathsf{else}}}
\newcommand{\CASE}{\operatorname{\mathsf{case}}}
\newcommand{\OF}{\operatorname{\mathsf{of}}}
\newcommand{\IN}{\operatorname{\mathsf{in}}}
\newcommand{\EMPTY}{\operatorname{\mathbf{0}}}
\newcommand{\ONE}{\operatorname{\mathbf{1}}}
\newcommand{\APP}{\circledast}
\newcommand{\NAT}{\operatorname{\mathbf{N}}}
\newcommand{\ZERO}{\operatorname{\mathsf{zero}}}
\newcommand{\SUCC}{\operatorname{\mathsf{succ}}}
\newcommand{\BOX}{\operatorname{\mathsf{box}}}
\newcommand{\WITH}{\operatorname{\mathsf{with}}}
\newcommand{\PREV}{\operatorname{\mathsf{prev}}}
\newcommand{\UNBOX}{\operatorname{\mathsf{unbox}}}
\newcommand{\BOXSUM}{\operatorname{\mathsf{box}^+}}
\newcommand{\OPNAME}{\operatorname{\mathsf{op}}}

\newcommand{\Exp}{\operatorname{Exp}}
\newcommand{\ClType}{\operatorname{ClType}}
\newcommand{\red}{\mathrel{\mapsto}}
\newcommand{\redrt}{\mathrel{\rightsquigarrow}}
\newcommand{\llrr}[1]{\llbracket #1 \rrbracket}
\newcommand{\defeq}{\triangleq}
\newcommand{\bnfeq}{\mathrel{::=}}
\newcommand{\trees}{\mathcal{S}}
\newcommand{\res}[2]{r^{#1}_{#2}}
\newcommand{\den}[1]{\llbracket#1\rrbracket}
\newcommand{\lrel}[2]{R^{#2}_{#1}}
\newcommand{\boxd}{\mathsf{bd}}
\newcommand{\usize}{\mathsf{us}}
\newcommand{\natto}{\mathrel{\dot{\to}}}
\newcommand{\ceq}{\simeq_{\mathsf{ctx}}}
\newcommand{\Sub}[1]{\mathsf{Sub}(#1)}
\newcommand{\height}[2]{\mathsf{height}_{#2}(#1)}

\newcommand{\ThetaLetter}{\mathchar"7002}
\renewcommand{\Theta}{\operatorname{\mathsf{fix}}}
\newcommand{\Rec}{\operatorname{\mathsf{Rec}}}

\newcommand{\explsubst}{\leftarrow}

\newcommand{\tinylater}{\scriptscriptstyle\LATER}

\newcommand{\guarded}[1]{\ensuremath{#1^{\mathsf{g}}}}

\newcommand{\gStream}[1]{\guarded{\mathsf{Str}}#1}
\newcommand{\Stream}[1]{\mathsf{Str}#1}
\newcommand{\head}{\operatorname{\guarded{\mathsf{hd}}}}
\newcommand{\tail}{\operatorname{\guarded{\mathsf{tl}}}}
\newcommand{\limhead}{\operatorname{\mathsf{hd}}}
\newcommand{\limtail}{\operatorname{\mathsf{tl}}}

\newcommand{\gCoNat}{\guarded{\mathsf{CoNat}}}
\newcommand{\CoNat}{\mathsf{CoNat}}
\newcommand{\gpred}{\operatorname{\guarded{\mathsf{pred}}}}
\newcommand{\pred}{\operatorname{\mathsf{pred}}}

\newcommand{\iterate}{\operatorname{\mathsf{iterate}}}
\newcommand{\cons}{\operatorname{\mathsf{cons}}}
\newcommand{\consin}{\mathbin{::}}
\newcommand{\gsecond}{\guarded{\mathsf{2nd}}}
\newcommand{\gthird}{\guarded{\mathsf{3rd}}}
\newcommand{\decons}{\operatorname{\mathsf{decons}}}
\newcommand{\naturals}{\operatorname{\mathsf{nats}}}
\renewcommand{\interleave}{\operatorname{\mathsf{interleave}}}
\newcommand{\tyrol}{\operatorname{\mathsf{toggle}}}
\newcommand{\folds}{\operatorname{\mathsf{paperfolds}}}
\newcommand{\zeros}{\operatorname{\mathsf{zeros}}}
\newcommand{\limit}{\operatorname{\mathsf{lim}}}
\newcommand{\limittwo}{\operatorname{\mathsf{lim2}}}
\renewcommand{\lim}{\limit}
\newcommand{\everysecond}{\operatorname{\mathsf{every2nd}}}
\newcommand{\plus}{\operatorname{\guarded{\mathsf{plus}}}}
\newcommand{\map}{\operatorname{\guarded{\mathsf{map}}}}
\newcommand{\constmap}{\operatorname{\mathsf{map}}}
\newcommand{\limplus}{\operatorname{\mathsf{plus}}}

\newcommand{\hastype}[3]{\ensuremath{{#1 \vdash #2 : #3}}}

\newcommand{\eps}{\varepsilon}

\newcommand{\El}{\ensuremath{\mathcal{E}}}
\newcommand{\Sl}{\ensuremath{\mathcal{S}}}
\newcommand{\Ul}{\ensuremath{\mathcal{U}}}
\newcommand{\Dl}{\ensuremath{\mathcal{D}}}
\newcommand{\Fl}{\ensuremath{\mathcal{F}}}
\newcommand{\Pl}{\ensuremath{\mathcal{P}}}
\newcommand{\Tl}{\ensuremath{\mathcal{T}}}
\newcommand{\Ml}{\ensuremath{\mathcal{M}}}
\newcommand{\Il}{\ensuremath{\mathcal{I}}}
\newcommand{\Cl}{\ensuremath{\mathcal{C}}}
\newcommand{\Bl}{\ensuremath{\mathcal{B}}}
\newcommand{\Al}{\ensuremath{\mathcal{A}}}
\newcommand{\Gl}{\ensuremath{\mathcal{G}}}
\newcommand{\Nl}{\ensuremath{\mathcal{N}}}
\newcommand{\BB}{\ensuremath{\mathbb{B}}}
\newcommand{\CC}{\ensuremath{\mathbb{C}}}
\newcommand{\KK}{\ensuremath{\mathbb{K}}}
\newcommand{\NN}{\NAT}
\newcommand{\PP}{\ensuremath{\mathbb{P}}}
\newcommand{\VV}{\ensuremath{\mathbb{V}}}
\newcommand{\UU}{\ensuremath{\mathbb{U}}}
\newcommand{\DD}{\ensuremath{\mathbb{D}}}
\newcommand{\EE}{\ensuremath{\mathbb{E}}}
\newcommand{\TT}{\ensuremath{\mathbb{T}}}

\newcommand{\comp}{\circ}

\newcommand{\id}[1]{\ensuremath{\text{id}_{#1}}}
\newcommand{\inv}[1]{\ensuremath{#1^{-1}}}
\newcommand{\iso}{\cong}
\newcommand{\isetsep}{\ensuremath{{\,\middle|\,}}}

\newcommand{\later}{\operatorname\triangleright}
\newcommand{\always}{\operatorname\square}
\newcommand{\lift}{\operatorname{\mathsf{lift}}}
\newcommand{\liftstr}[1]{#1_{\guarded{\mathsf{Str}}}}

\newcommand{\eqlaternextrule}{\ensuremath{\textsc{eq}^{\later}_{\NEXT}}}

\renewcommand{\implies}{\Rightarrow}
\renewcommand{\iff}{\Leftrightarrow}

\newcommand{\inhab}[1]{\ensuremath{\mathrm{Inhab}\left(#1\right)}}
\newcommand{\total}[1]{\ensuremath{\mathrm{TI}\left(#1\right)}}

\newcommand{\defined}{\ensuremath{\overset{\triangle}{=}}}

\newcommand{\op}[1]{\ensuremath{#1^{\text{op}}}}
\newcommand{\CAT}{\ensuremath{\mathbf{Cat}}}
\newcommand{\sets}{\ensuremath{\mathbf{Set}}}
\newcommand{\Sh}[1]{\ensuremath{\text{Sh}\left(#1\right)}}
\newcommand{\PSh}[1]{\ensuremath{\text{PSh}\left(#1\right)}}
\newcommand{\subobj}[1]{\ensuremath{\mathbf{Sub}\left(#1\right)}}
\renewcommand{\hom}[3]{\ensuremath{\text{Hom}_{#1}\left(#2,#3\right)}}
\newcommand{\nxt}{\ensuremath{\mathbf{next}}}
\renewcommand{\restriction}{\mathord{\upharpoonright}}

\newenvironment{diagram}{\begin{tikzcd}[row sep=1.5cm,column sep=1.5cm]}{\end{tikzcd}}
\newenvironment{largediagram}{\begin{tikzcd}[row sep=2.6cm,column sep=2.6cm]}{\end{tikzcd}}
\newenvironment{smalldiagram}{\begin{tikzcd}[row sep=1cm,column sep=1cm]}{\end{tikzcd}}

\newcommand{\denS}[1]{\ensuremath{\left\llbracket #1 \right\rrbracket_{\Sl}}}
\newcommand{\denSet}[1]{\ensuremath{\left\llbracket #1 \right\rrbracket_{\sets}}}
